\documentclass[aps,preprint,tightenlines,superscriptaddress,showpacs]{revtex4-1}
\usepackage{times}
\usepackage{amsmath}
\usepackage{graphicx}
\usepackage{epsfig}
\usepackage{dcolumn}
\usepackage{bm}
\usepackage{multirow}
\usepackage{url}
\usepackage[colorlinks,linkcolor=red,anchorcolor=green,citecolor=blue]{hyperref}

\newcommand{\BR}{{\cal B}}

\newcommand{\beq}{\begin{equation}}
\newcommand{\eeq}{\end{equation}}
\newcommand{\bitm}{\begin{itemize}}
\newcommand{\eitm}{\end{itemize}}

\begin{document}


\title{
\boldmath
Evidence of $\Upsilon(1S)\rightarrow J/\psi+\chi_{c1}$
and search for double-charmonium production in $\Upsilon(1S)$ and $\Upsilon(2S)$ decays}


\noaffiliation
\affiliation{University of the Basque Country UPV/EHU, 48080 Bilbao}
\affiliation{Beihang University, Beijing 100191}
\affiliation{University of Bonn, 53115 Bonn}
\affiliation{Budker Institute of Nuclear Physics SB RAS and Novosibirsk State University, Novosibirsk 630090}
\affiliation{Faculty of Mathematics and Physics, Charles University, 121 16 Prague}
\affiliation{Chonnam National University, Kwangju 660-701}
\affiliation{University of Cincinnati, Cincinnati, Ohio 45221}
\affiliation{Deutsches Elektronen--Synchrotron, 22607 Hamburg}
\affiliation{Justus-Liebig-Universit\"at Gie\ss{}en, 35392 Gie\ss{}en}
\affiliation{The Graduate University for Advanced Studies, Hayama 240-0193}
\affiliation{Gyeongsang National University, Chinju 660-701}
\affiliation{Hanyang University, Seoul 133-791}
\affiliation{University of Hawaii, Honolulu, Hawaii 96822}
\affiliation{High Energy Accelerator Research Organization (KEK), Tsukuba 305-0801}
\affiliation{IKERBASQUE, Basque Foundation for Science, 48011 Bilbao}
\affiliation{Indian Institute of Technology Guwahati, Assam 781039}
\affiliation{Indian Institute of Technology Madras, Chennai 600036}
\affiliation{Indiana University, Bloomington, Indiana 47408}
\affiliation{Institute of High Energy Physics, Chinese Academy of Sciences, Beijing 100049}
\affiliation{Institute of High Energy Physics, Vienna 1050}
\affiliation{Institute for High Energy Physics, Protvino 142281}
\affiliation{INFN - Sezione di Torino, 10125 Torino}
\affiliation{Institute for Theoretical and Experimental Physics, Moscow 117218}
\affiliation{J. Stefan Institute, 1000 Ljubljana}
\affiliation{Kanagawa University, Yokohama 221-8686}
\affiliation{Institut f\"ur Experimentelle Kernphysik, Karlsruher Institut f\"ur Technologie, 76131 Karlsruhe}
\affiliation{Kennesaw State University, Kennesaw GA 30144}
\affiliation{Department of Physics, Faculty of Science, King Abdulaziz University, Jeddah 21589}
\affiliation{Korea Institute of Science and Technology Information, Daejeon 305-806}
\affiliation{Korea University, Seoul 136-713}
\affiliation{Kyungpook National University, Daegu 702-701}
\affiliation{\'Ecole Polytechnique F\'ed\'erale de Lausanne (EPFL), Lausanne 1015}
\affiliation{Faculty of Mathematics and Physics, University of Ljubljana, 1000 Ljubljana}
\affiliation{Luther College, Decorah, Iowa 52101}
\affiliation{University of Maribor, 2000 Maribor}
\affiliation{Max-Planck-Institut f\"ur Physik, 80805 M\"unchen}
\affiliation{School of Physics, University of Melbourne, Victoria 3010}
\affiliation{Moscow Physical Engineering Institute, Moscow 115409}
\affiliation{Moscow Institute of Physics and Technology, Moscow Region 141700}
\affiliation{Graduate School of Science, Nagoya University, Nagoya 464-8602}
\affiliation{Kobayashi-Maskawa Institute, Nagoya University, Nagoya 464-8602}
\affiliation{Nara Women's University, Nara 630-8506}
\affiliation{National Central University, Chung-li 32054}
\affiliation{National United University, Miao Li 36003}
\affiliation{Department of Physics, National Taiwan University, Taipei 10617}
\affiliation{H. Niewodniczanski Institute of Nuclear Physics, Krakow 31-342}
\affiliation{Niigata University, Niigata 950-2181}
\affiliation{Osaka City University, Osaka 558-8585}
\affiliation{Pacific Northwest National Laboratory, Richland, Washington 99352}
\affiliation{Peking University, Beijing 100871}
\affiliation{University of Science and Technology of China, Hefei 230026}
\affiliation{Seoul National University, Seoul 151-742}
\affiliation{Soongsil University, Seoul 156-743}
\affiliation{Sungkyunkwan University, Suwon 440-746}
\affiliation{School of Physics, University of Sydney, NSW 2006}
\affiliation{Department of Physics, Faculty of Science, University of Tabuk, Tabuk 71451}
\affiliation{Tata Institute of Fundamental Research, Mumbai 400005}
\affiliation{Excellence Cluster Universe, Technische Universit\"at M\"unchen, 85748 Garching}
\affiliation{Tohoku University, Sendai 980-8578}
\affiliation{Department of Physics, University of Tokyo, Tokyo 113-0033}
\affiliation{Tokyo Institute of Technology, Tokyo 152-8550}
\affiliation{Tokyo Metropolitan University, Tokyo 192-0397}
\affiliation{University of Torino, 10124 Torino}
\affiliation{CNP, Virginia Polytechnic Institute and State University, Blacksburg, Virginia 24061}
\affiliation{Wayne State University, Detroit, Michigan 48202}
\affiliation{Yamagata University, Yamagata 990-8560}
\affiliation{Yonsei University, Seoul 120-749}
  \author{S.~D.~Yang}\affiliation{Peking University, Beijing 100871} 
  \author{C.~P.~Shen}\affiliation{Beihang University, Beijing 100191} 
  \author{Y.~Ban}\affiliation{Peking University, Beijing 100871} 
  \author{A.~Abdesselam}\affiliation{Department of Physics, Faculty of Science, University of Tabuk, Tabuk 71451} 
  \author{I.~Adachi}\affiliation{High Energy Accelerator Research Organization (KEK), Tsukuba 305-0801}\affiliation{The Graduate University for Advanced Studies, Hayama 240-0193} 
  \author{H.~Aihara}\affiliation{Department of Physics, University of Tokyo, Tokyo 113-0033} 
  \author{S.~Al~Said}\affiliation{Department of Physics, Faculty of Science, University of Tabuk, Tabuk 71451}\affiliation{Department of Physics, Faculty of Science, King Abdulaziz University, Jeddah 21589} 
  \author{K.~Arinstein}\affiliation{Budker Institute of Nuclear Physics SB RAS and Novosibirsk State University, Novosibirsk 630090} 
  \author{D.~M.~Asner}\affiliation{Pacific Northwest National Laboratory, Richland, Washington 99352} 
  \author{V.~Aulchenko}\affiliation{Budker Institute of Nuclear Physics SB RAS and Novosibirsk State University, Novosibirsk 630090} 
% ...
\collaboration{The Belle Collaboration}

\begin{abstract}
  Using data samples of $102\times10^6$ $\Upsilon(1S)$ and $158\times10^6$ $\Upsilon(2S)$ events collected with the Belle detector,
  a first experimental search has been made for double-charmonium production in the exclusive decays
  $\Upsilon(1S,2S)\rightarrow J/\psi(\psi')+X$, where $X=\eta_c$, $\chi_{cJ} (J=~0,~1,~2)$, $\eta_c(2S)$, $X(3940)$,
  and $X(4160)$. No significant signal is observed in the spectra of the mass recoiling against the reconstructed $J/\psi$ or $\psi'$
  except for the evidence of $\chi_{c1}$ production with a significance of $4.6\sigma$
  for $\Upsilon(1S)\rightarrow J/\psi+\chi_{c1}$.
  The measured branching fraction $\BR(\Upsilon(1S)\rightarrow J/\psi+\chi_{c1})$ is $(3.90\pm1.21(\rm stat.)\pm0.23 (\rm syst.))\times10^{-6}$.
  The $90\%$ confidence level upper limits on the branching fractions
  of the other modes having a significance of less than $3\sigma$ are determined.
  These results are consistent with theoretical calculations using the nonrelativistic QCD factorization approach.
\end{abstract}


\maketitle


    For many years, one of the largest discrepancies in quarkonium physics
    has been the unexpected disagreement between the experimental measurements and theoretical predictions
    for double-charmonium production at {\it B} factories.
    The cross sections of the processes $e^+e^- \rightarrow J/\psi\eta_c$, $J/\psi\eta_c(2S)$,
    $\psi^\prime\eta_c$, $\psi^\prime\eta_c(2S)$, $J/\psi\chi_{c0}$, and $\psi^\prime\chi_{c0}$
    measured by the Belle \cite{PhysRevLett.89.142001,PhysRevD.70.071102}
    and BaBar \cite{PhysRevD.72.031101} Collaborations
    exceeded the leading-order non-relativistic QCD (NRQCD) calculations by approximately an order of magnitude
    \cite{PhysRevD.67.054007,*PhysRevD.72.099901,PhysLettB.557.45,PhysRevD.70.074007,PhysRevD.69.094027,PhysLettB.612.215,PhysRevD.77.014002}.
    Later the double-charmonium productions of $J/\psi X(3940)$ \cite{PhysRevLett.98.082001,PhysRevD.79.071101}
    and $J/\psi X(4160)$ \cite{PhysRevLett.100.202001} were observed in $e^+e^-$ annihilation by Belle as well.
    Numerous theoretical investigations in the following years had attempted to alleviate
    this disquieting discrepancy, and it is now believed that one can achieve agreement within reasonable uncertainties
    when both the QCD radiative and
    relativistic corrections of the order of $\upsilon^2$ (where $\upsilon$ is the quark relative velocity) are taken into account
    \cite{PhysRevLett.90.162001,PhysRevLett.96.092001,PhysRevD.75.074011,PhysRevD.77.094017,PhysRevD.77.094018,PhysRevD.79.074018,PhysRevD.85.114018,EPJC.71.1534,hep-ph1404.3723}.

    Inspired by the unexpectedly high double-charmonium production in $e^+e^-$ annihilation, interest has turned to
    the double-charmonium states produced in bottomonium decays.
    Several theoretical calculations have focused on these processes in perturbative QCD, \textit{e.g.,}
    $\eta_b\rightarrow J/\psi J/\psi$ \cite{PhysRevD.78.054003,PhysLettB.670.350,PhysRevD.81.014012,PhysLettB.702.49}
    and $\chi_{b0,1,2}\rightarrow J/\psi J/\psi$
    \cite{PhysRevD.72.094018,PhysRevD.80.094008,*PhysRevD.85.119901,PhysRevD.84.094031,JHEP2012.168}.
    Experimentally, however, such studies are extremely sparse,
    apart from the recent searches for several channels of $\chi_{bJ}$ into double charmonia
    for the first time by the Belle Collaboration \cite{PhysRevD.85.071102}.
    The measurements are consistent with NRQCD predictions, though no significant signals are observed;
    it is reasonable to extend the search for double-charmonium production to the C-odd $\Upsilon$ decays.
    Compared with the $\Upsilon(4S)$ resonance with its rather broad width,
    the first three $\Upsilon$ resonances are so narrow
    that the resonant decay contributions dominate over the continuum ones.
    This provides a further opportunity to probe
    the potential properties of double-charmonium production
    at these $\Upsilon$ peaks.

    Comprehensive studies of the exclusive decay of $\Upsilon$ into
    a vector-plus-pseudoscalar charmonium \cite{PhysRevD.76.074007},
    as well as the {\it S}-wave charmonium $J/\psi$ plus the {\it P}-wave charmonium $\chi_{cJ}$ $(J=0,~1,~2)$ \cite{PhysRevD.87.094004}
    have been performed in the NRQCD factorization approach,
    where the contributions from the strong, electromagnetic, and radiative channels
    were considered and the strong decay was taken as dominant.
    The branching fractions are predicted to be of order $10^{-6}$
    for $\Upsilon(nS)\rightarrow J/\psi(\psi^\prime)+\eta_c(\eta_c(2S))$ $(n=1,~2,~3)$ \cite{PhysRevD.76.074007};
    for $\Upsilon\rightarrow J/\psi+\eta_c$, in particular, the predicted branching fraction
    is consistent with the previous calculation of $1.7\times10^{-6}$
    with only the three-gluon contribution considered~\cite{PhysRevD.42.1577}.
    For the $J/\psi+\chi_{c0,1,2}$ decay modes,
    the branching fractions are calculated at the lowest order \cite{PhysRevD.87.094004};
    that of $\Upsilon(nS)\rightarrow J/\psi+\chi_{c1}$ is the
    largest---of order $10^{-6}$---while that of $J/\psi+\chi_{c2}$ is only of order $10^{-7}$.
    In this paper, we report studies of exclusive hadronic decays of $\Upsilon(1S)$ and $\Upsilon(2S)$ resonances
    to the double-charmonium final states $J/\psi(\psi')+X$, where $X$ is one of the $\eta_c$, $\chi_{cJ}~(J=0,~1,~2)$, $\eta_c(2S)$,
    $X(3940)$, and $X(4160)$ states.
    To improve the signal detection efficiencies,
    only the $J/\psi$ or $\psi'$ candidate is fully reconstructed;
    we search for the other charmonium state $X$ in the recoil mass distribution of
     the fully reconstructed $J/\psi$ or $\psi'$ candidate.
    The recoil mass is calculated as $M_{\mathrm{recoil}}(c\bar{c}) = \sqrt{(E_{\mathrm{CM}}-E^*_{c\bar{c}})^2-p_{c\bar{c}}^{*2}c^2}/c^2$,
    where $c\bar{c}$ is the reconstructed charmonium $J/\psi$ or $\psi'$,
    $E^*_{c\bar{c}}$ and $p^*_{c\bar{c}}$ are the center-of-mass (CM) energy and momentum of $J/\psi(\psi')$
    and $E_{\mathrm{CM}}$ is the center-of-mass energy of the colliding $e^+e^-$ system.


    This analysis utilizes the $\Upsilon(1S)$ and $\Upsilon(2S)$ samples from Belle
    with a total luminosity of $5.74$~fb$^{-1}$ ($102\times10^6$ events)
    and $24.91$~fb$^{-1}$ ($158\times10^6$ events), respectively.
    A $89.45$~fb$^{-1}$ data sample collected at $\sqrt{s}=10.52~\mathrm{GeV}$ is used
    to estimate the possible irreducible continuum contributions.
    All data were collected with the Belle detector \cite{Abashian2002117,PTEP201204D001}
    operating at the KEKB asymmetric-energy $e^{+}e^{-}$ collider \cite{Kurokawa20031,PTEP201303A001}.
    The signal Monte Carlo (MC) events are generated with EVTGEN \cite{Lange2001152}
    using the helicity-amplitude model \cite{PhysRevD.76.074007,PhysRevD.87.094004}.
    The decays of the two charmonium daughters are generated according to the
    known branching fractions~\cite{PhysRevD.86.010001},
    while unknown decay channels are generated by the Lund fragmentation model in PYTHIA \cite{JHEP2006.026}.
    Generic decay samples of $\Upsilon(1S)$ and $\Upsilon(2S)$ MC events
    produced using PYTHIA \cite{JHEP2006.026} with four times the luminosity of the data,
    are used to identify possible peaking backgrounds from $\Upsilon(1S)$ and $\Upsilon(2S)$ decays.

    The Belle detector is a large solid angle magnetic spectrometer that consists of a silicon vertex detector (SVD), a 50-layer central drift chamber (CDC), an array of aerogel threshold Cherenkov counters (ACC), a barrel-like arrangement of time-of-flight scintillation counters (TOF), and an electromagnetic calorimeter comprising CsI(Tl) crystals located inside a superconducting solenoid coil that provides a 1.5 T magnetic field. An iron flux return located outside the coil is instrumented to detect $K^{0}_{L}$ mesons and to identify muons.  A detailed description of the Belle detector can be found
    in Ref.~\cite{Abashian2002117}.


    Primary charged tracks are selected with $dr<2~\mathrm{cm}$ and $|dz|<4~\mathrm{cm}$,
    where $dr$ and $dz$ are the impact parameters perpendicular to
    and along the beam direction with respect to the interaction point.
    In addition, the transverse momentum of every charged track in the laboratory frame
    is restricted to be larger than $0.1~\mathrm{GeV/{\mathit c}}$.
    QED backgrounds are significantly suppressed by the requirement
    that the charged multiplicity ($N_{\rm ch}$) in every event satisfies $N_{\rm ch}>4$ \cite{PhysRevD.70.071102}.
    Lepton candidate tracks from $J/\psi(\psi')$ are required to have a muon likelihood ratio
    $R_{\mu}=\frac{\mathcal{L}_{\mu}}{\mathcal{L}_{\mu}+\mathcal{L}_{K}+\mathcal{L}_{\pi}}>0.1$ \cite{Abashian49169}
    or an electron likelihood ratio $R_e=\frac{\mathcal{L}_e}{\mathcal{L}_e+\mathcal{L}_{non-e}}>0.01$ \cite{Hanagaki485490}.
    To reduce the effect of bremsstrahlung and final-state radiation,
    photons detected in the ECL within a $50~\mathrm{mrad}$ cone of the original electron or positron direction are included
    in the calculation of the $e^+/e^-$ four-momentum.
    The lepton-identification efficiencies for $e^\pm$ and $\mu^\pm$ are about $98\%$ and $96\%$, respectively.
    Because $\psi'$ is also reconstructed from $J/\psi\pi^+\pi^-$,
    charged tracks with $R_{K}=\frac{\mathcal{L}_{K}}{\mathcal{L}_K+\mathcal{L}_\pi}<0.4$ \cite{Nakano494402} are considered to be pions
    for this purpose, with an efficiency of about $98\%$ and a kaon misidentification rate of about $2.6\%$.

\begin{figure}[!htb]
      \centering
        \begin{minipage}[c]{0.33\textwidth}
        \centering
        \includegraphics[width=0.99\textwidth]{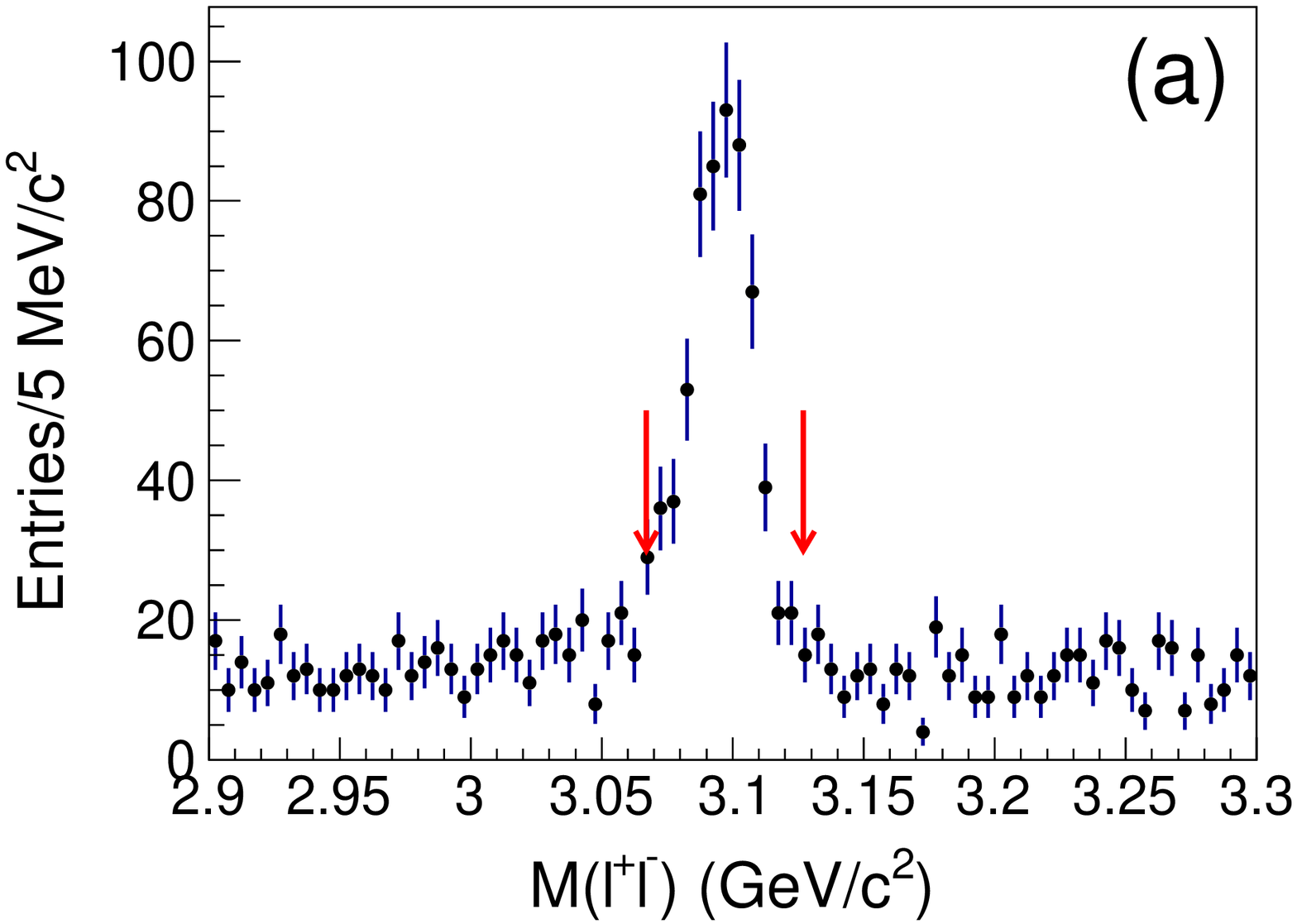}
        \end{minipage}%
        \begin{minipage}[c]{0.33\textwidth}
        \centering
        \includegraphics[width=0.99\textwidth]{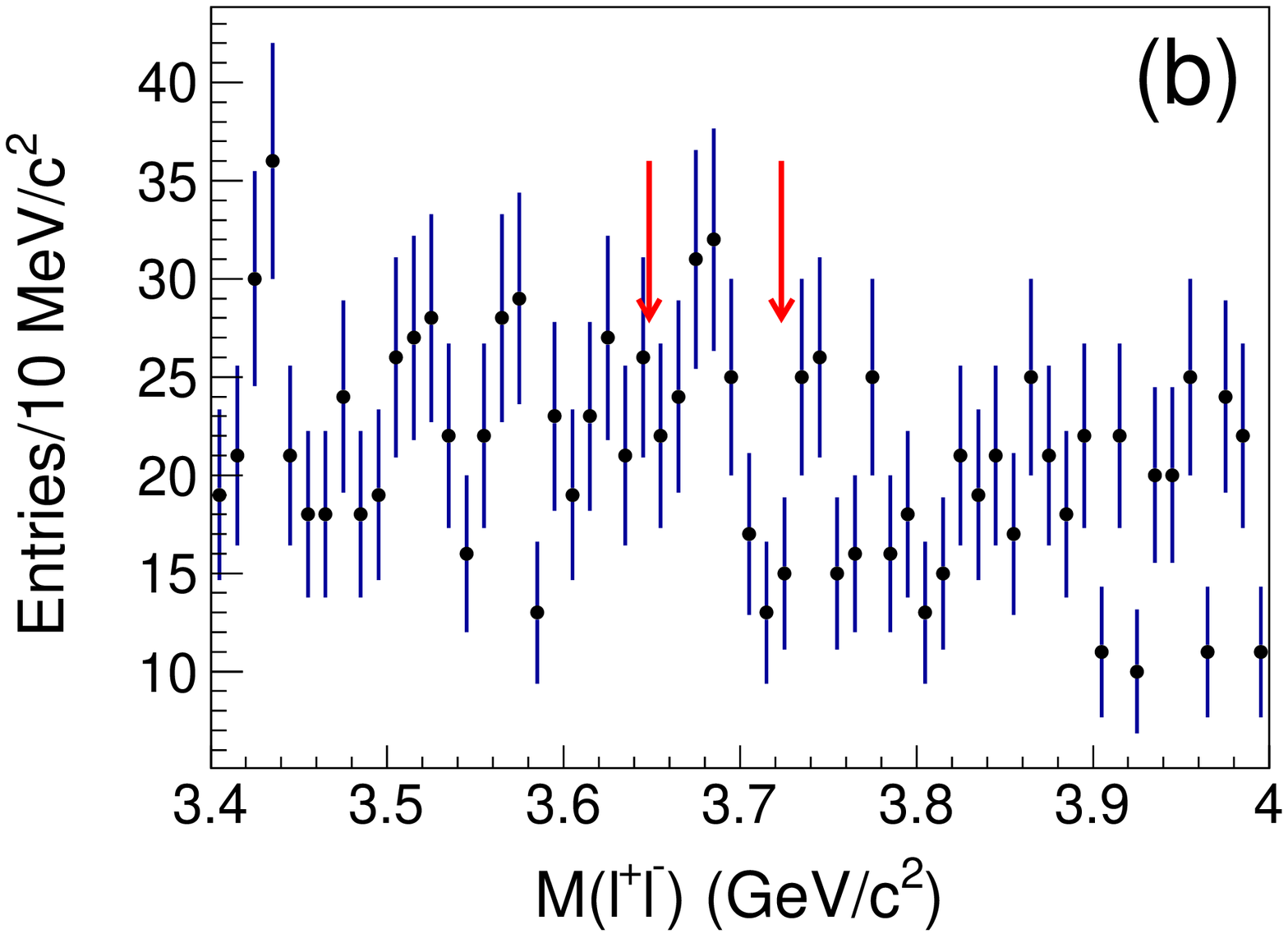}
        \end{minipage}%
        \begin{minipage}[c]{0.33\textwidth}
        \centering
        \includegraphics[width=0.99\textwidth]{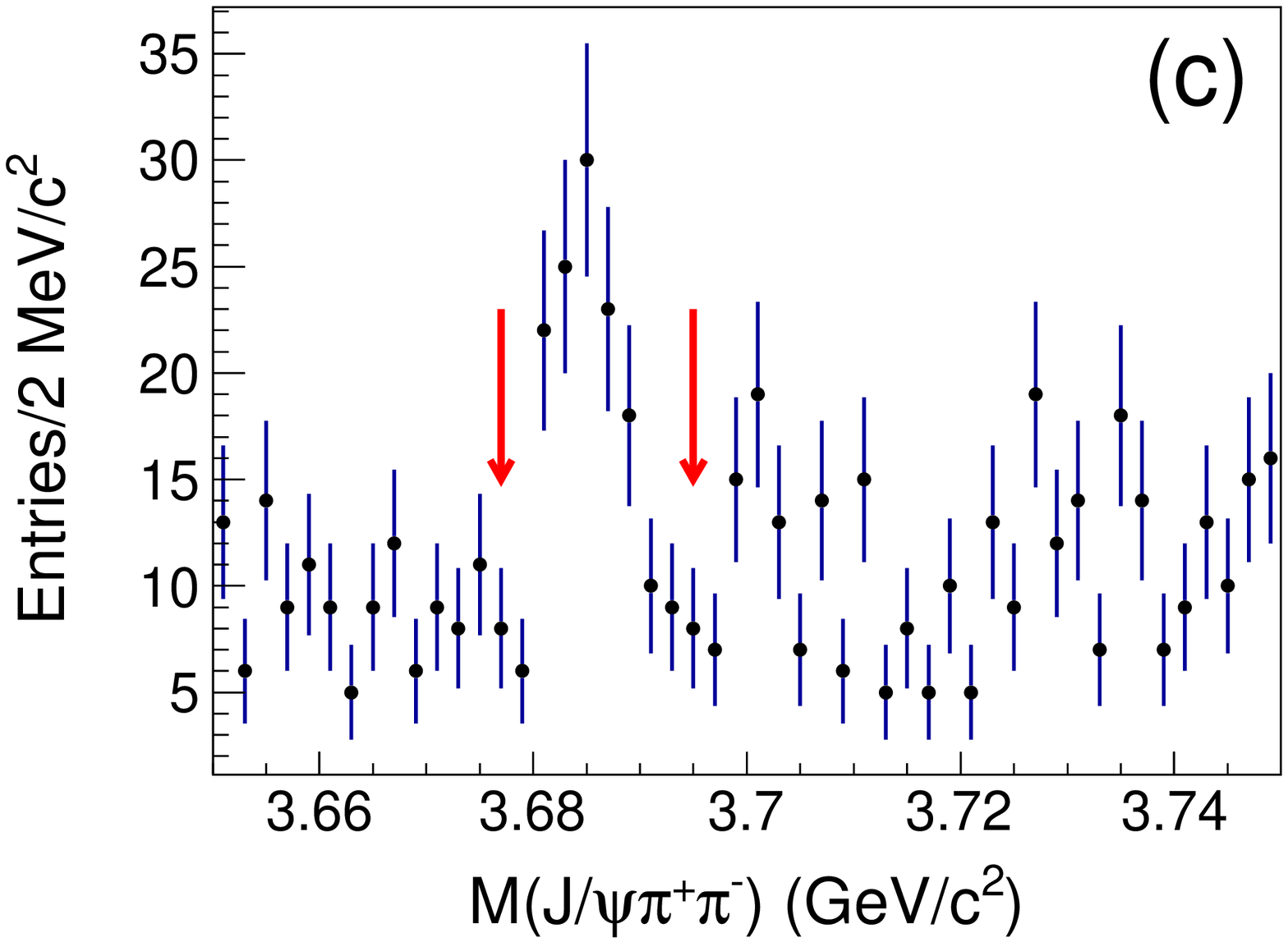}
        \end{minipage}\\
        \begin{minipage}[c]{0.33\textwidth}
        \centering
        \includegraphics[width=0.99\textwidth]{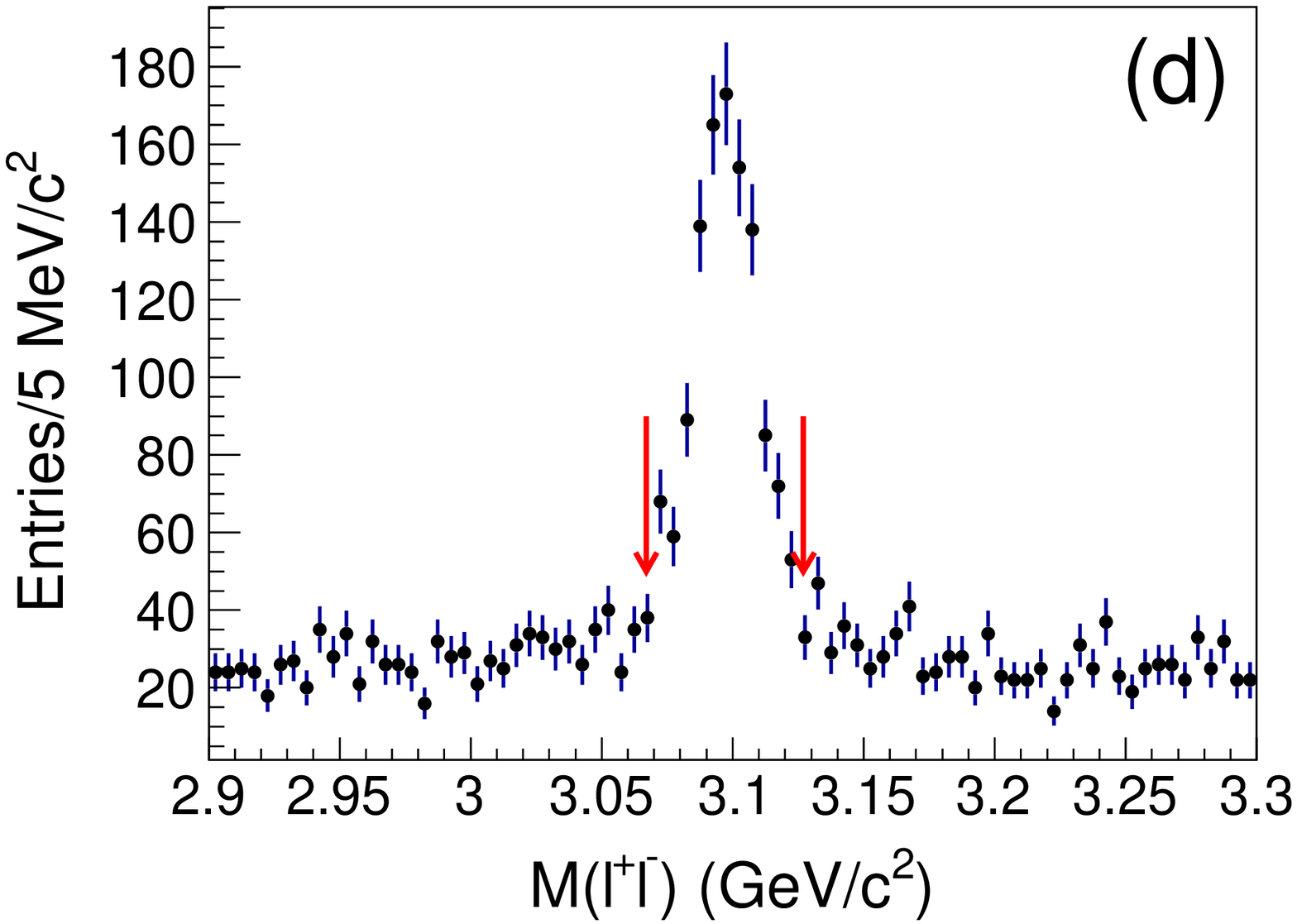}
        \end{minipage}%
        \begin{minipage}[c]{0.33\textwidth}
        \centering
        \includegraphics[width=0.99\textwidth]{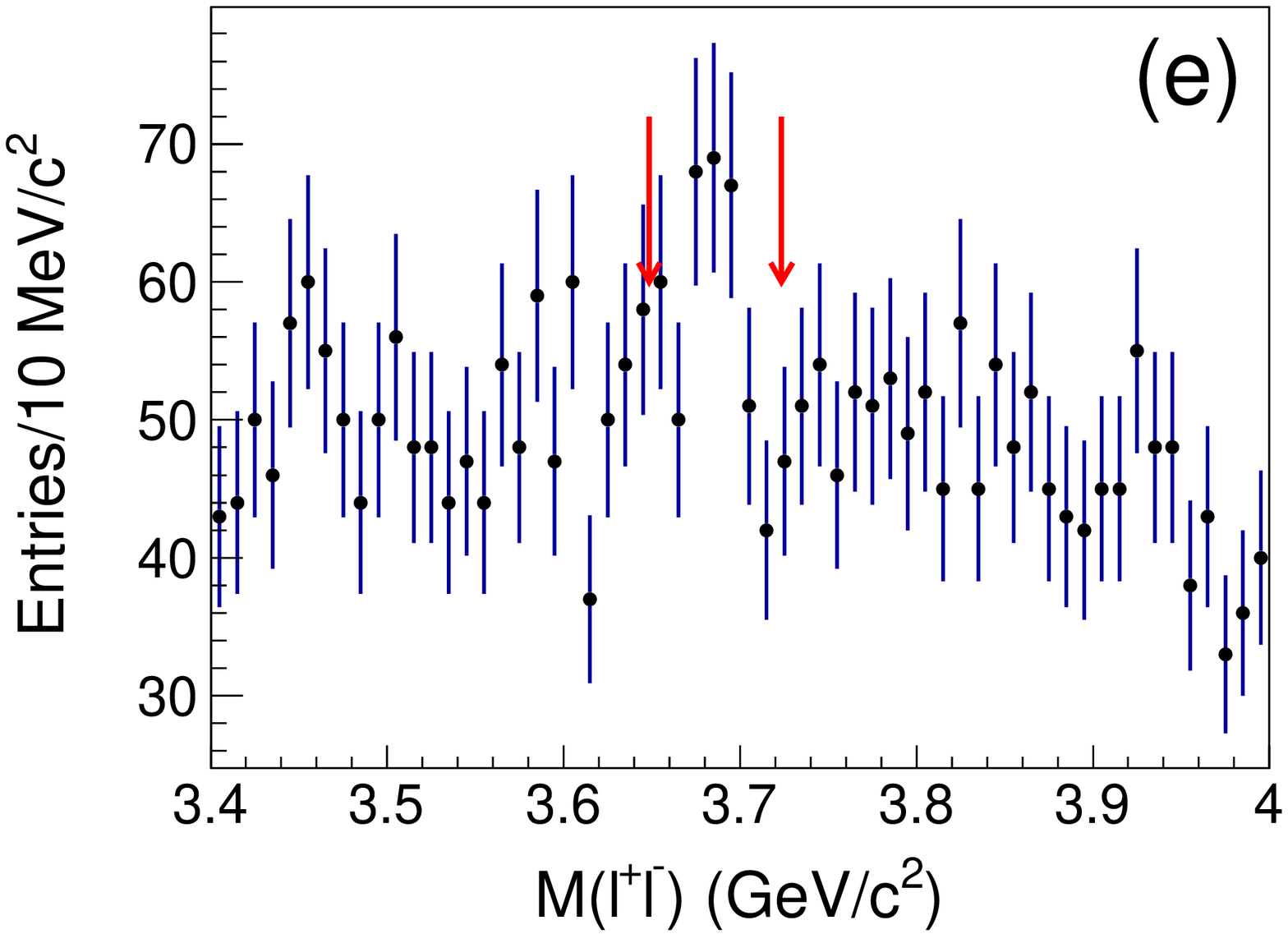}
        \end{minipage}%
        \begin{minipage}[c]{0.33\textwidth}
        \centering
        \includegraphics[width=0.99\textwidth]{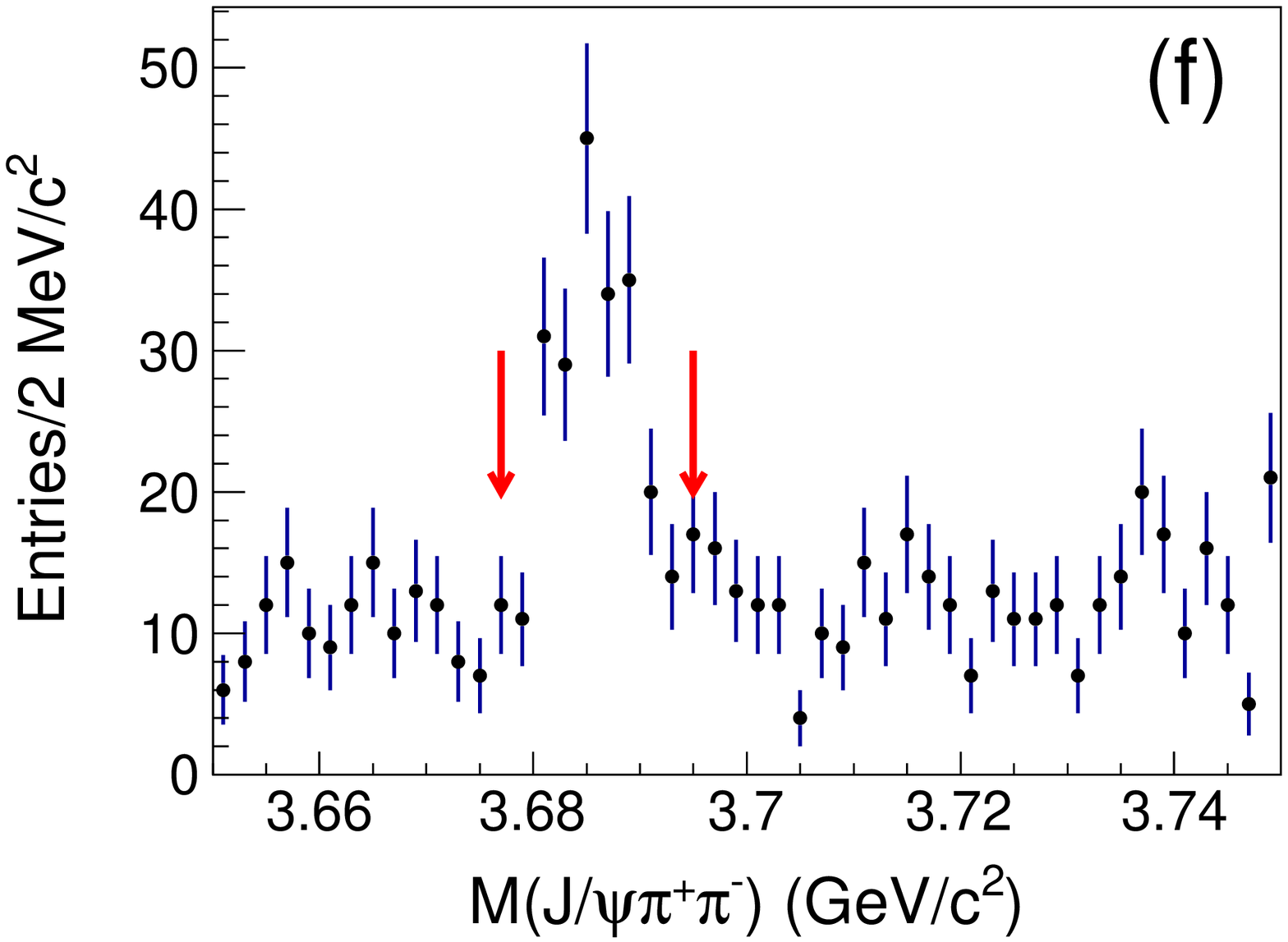}
        \end{minipage}
      \caption{{Distributions of the invariant masses of $J/\psi$ identified with lepton pairs,
      and $\psi'$ identified with both $\ell^+\ell^-$ and $J/\psi\pi^+\pi^-$
      with the $J/\psi(\psi')$ recoil mass within $2.2~\mathrm{GeV/{\mathit c}^2}$ and $4.6~\mathrm{GeV/{\mathit c}^2}$ from left to right.
      The upper and lower three graphs are for $\Upsilon(1S)$ and $\Upsilon(2S)$ decays, respectively.
      The arrows show the signal regions of $J/\psi$ or $\psi'$ masses.}}\label{fig-data-Y1S}
    \end{figure}

    When reconstructing $J/\psi(\psi')$ candidates for all the modes,
    a mass-constrained fit is applied 
    to improve the resolutions of the recoil mass distributions.
    MC simulations indicate that the $J/\psi(\psi')$ has almost
    the same mass resolution if the $J/\psi(\psi')$ is reconstructed from the same final states
    in $\Upsilon(1S,2S)\to J/\psi(\psi') + X$ processes.
    The signal region for $J/\psi$ is defined as
    $|M_{\ell^+\ell^-}-m_{J/\psi}|<0.03~\mathrm{GeV/{\mathit c}^2}~(\sim2.5\sigma)$, where $\ell=e$ or $\mu$
    and $m_{J/\psi}$ is the nominal mass of the $J/\psi$ \cite{PhysRevD.86.010001};
    the $J/\psi$ mass sidebands are defined as $2.97~\mathrm{GeV/{\mathit c}^2}<M_{\ell^+\ell^-}<3.03~\mathrm{GeV/{\mathit c}^2}$
    or $3.17~\mathrm{GeV/{\mathit c}^2}<M_{\ell^+\ell^-}<3.23~\mathrm{GeV/{\mathit c}^2}$.
    For $\psi'$ candidates with $\ell^+\ell^-$ and $J/\psi\pi^+\pi^-$ final states,
    the $\psi'$ signal regions are defined as $|M_{\ell^+\ell^-}-m_{\psi'}|<0.0375~\mathrm{GeV/{\mathit c}^2}~(\sim2.5\sigma)$
    and $|M_{J/\psi\pi^+\pi^-}-m_{\psi'}|<0.009~\mathrm{GeV/{\mathit c}^2}~(\sim3.0\sigma)$, respectively,
    where $m_{\psi'}$ is the nominal mass of the $\psi'$ \cite{PhysRevD.86.010001}.
    The $\psi'$ mass sidebands are defined as $3.535~\mathrm{GeV/{\mathit c}^2}<M_{\ell^+\ell^-}<3.610~\mathrm{GeV/{\mathit c}^2}$
    or $3.760~\mathrm{GeV/{\mathit c}^2}<M_{\ell^+\ell^-}<3.835~\mathrm{GeV/{\mathit c}^2}$,
    and $3.652~\mathrm{GeV/{\mathit c}^2}<M_{J/\psi\pi^+\pi^-}<3.670~\mathrm{GeV/{\mathit c}^2}$
    or $3.700~\mathrm{GeV/{\mathit c}^2}<M_{J/\psi\pi^+\pi^-}<3.718~\mathrm{GeV/{\mathit c}^2}$.
    The mass sidebands of both the $J/\psi$ and $\psi'$ are twice as wide as the signal region.
    Figure~\ref{fig-data-Y1S} shows the mass distributions of
    the reconstructed $J/\psi(\rightarrow \ell^+\ell^-)$, $\psi'(\rightarrow \ell^+\ell^-)$, and $\psi'(\rightarrow J/\psi\pi^+\pi^-)$ candidates
    in $\Upsilon(1S)$ and $\Upsilon(2S)$ decays.
    The signal regions of the $J/\psi$ and $\psi'$ candidates are indicated with arrows in the corresponding graphs.
    The analysis region of the recoil masses is
    $2.2~\mathrm{GeV/{\mathit c}^2}<M_{\mathrm{recoil}}(J/\psi(\psi'))<4.6~\mathrm{GeV/{\mathit c}^2}$
    and covers all the recoil charmonium states of interest.

    \begin{figure}[!htb]
      \centering
        \begin{minipage}[c]{0.33\textwidth}
        \centering
        \includegraphics[width=0.99\textwidth]{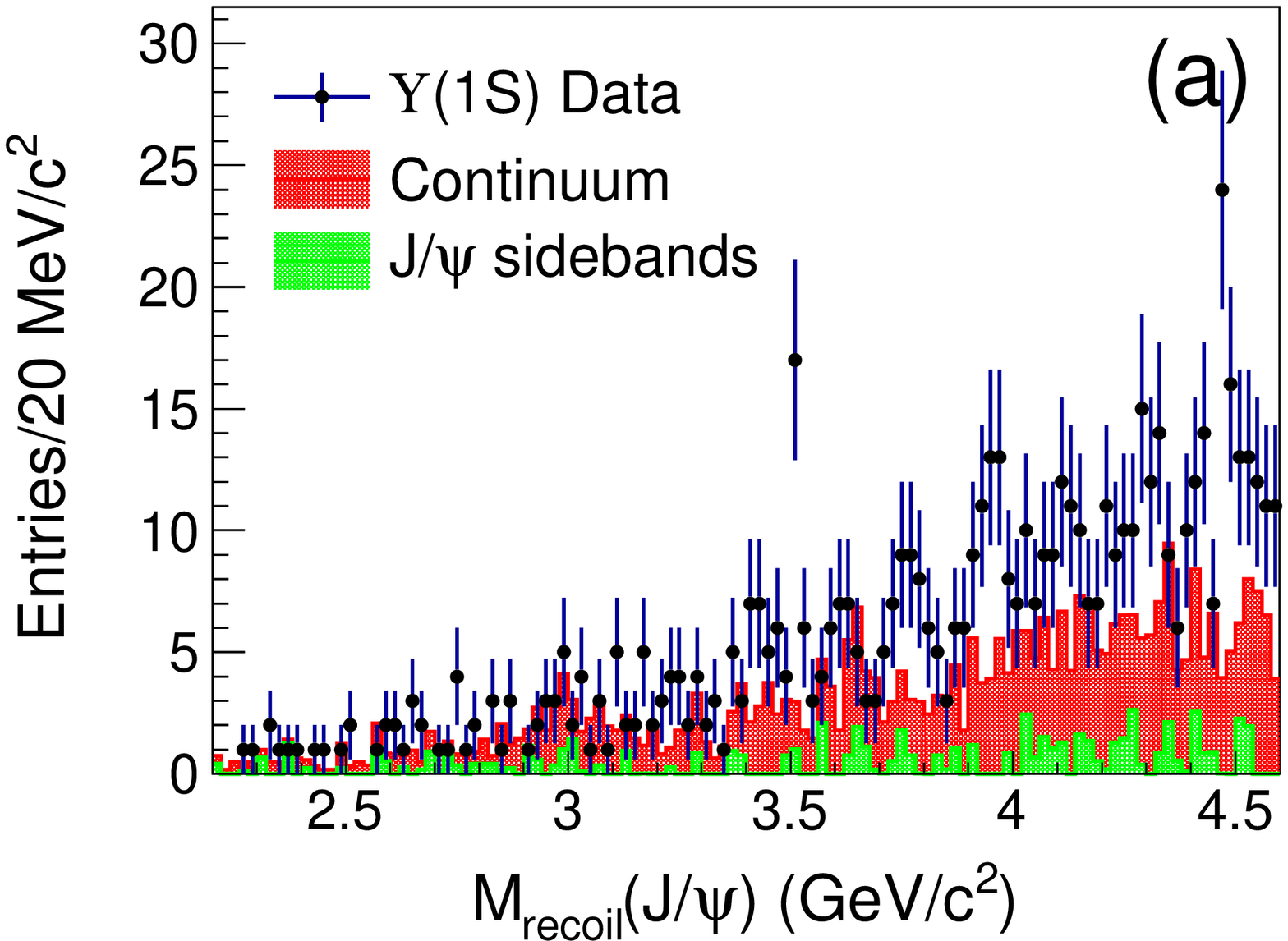}
        \end{minipage}%
        \begin{minipage}[c]{0.33\textwidth}
        \centering
        \includegraphics[width=0.99\textwidth]{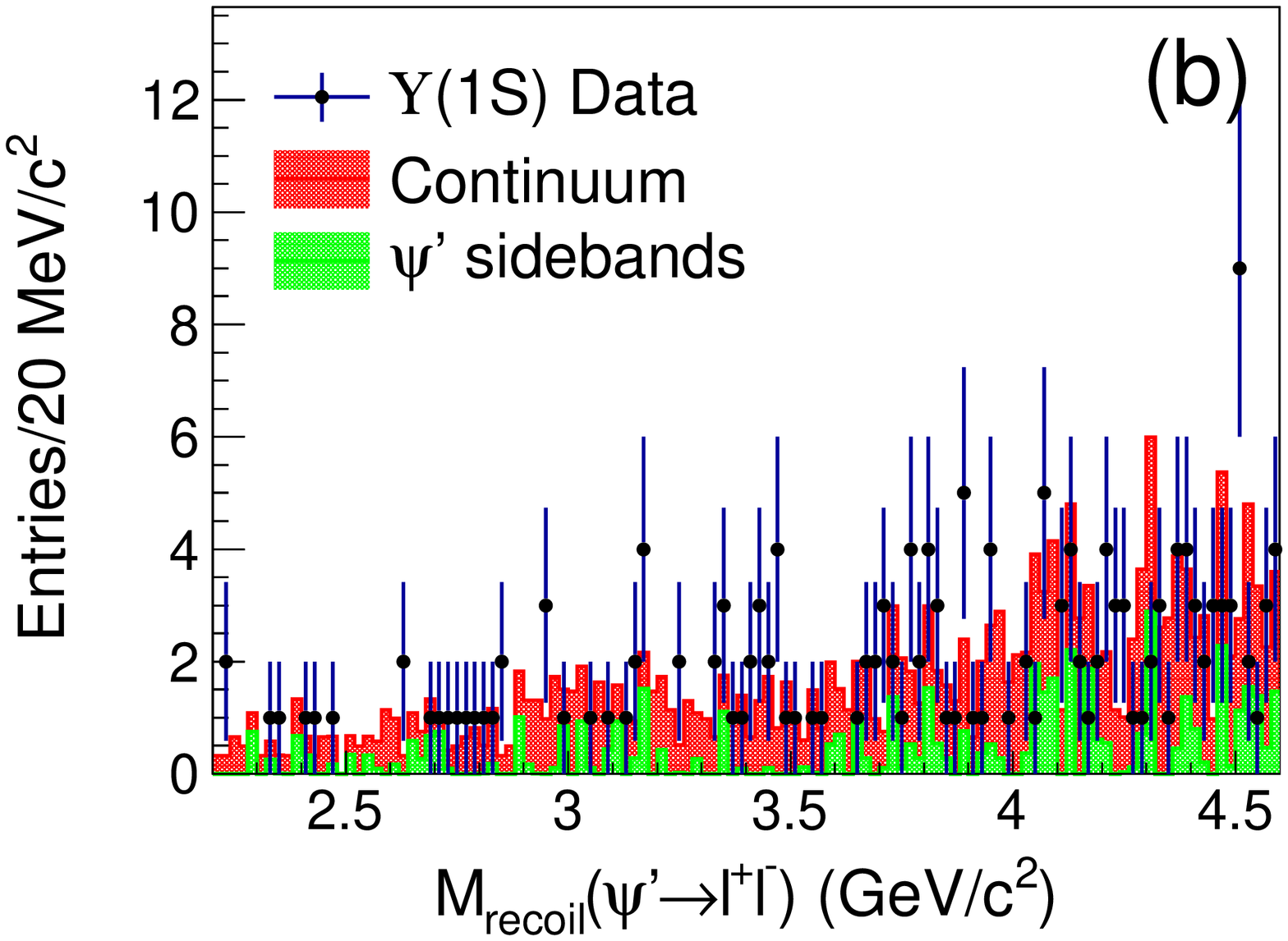}
        \end{minipage}%
        \begin{minipage}[c]{0.33\textwidth}
        \centering
        \includegraphics[width=0.99\textwidth]{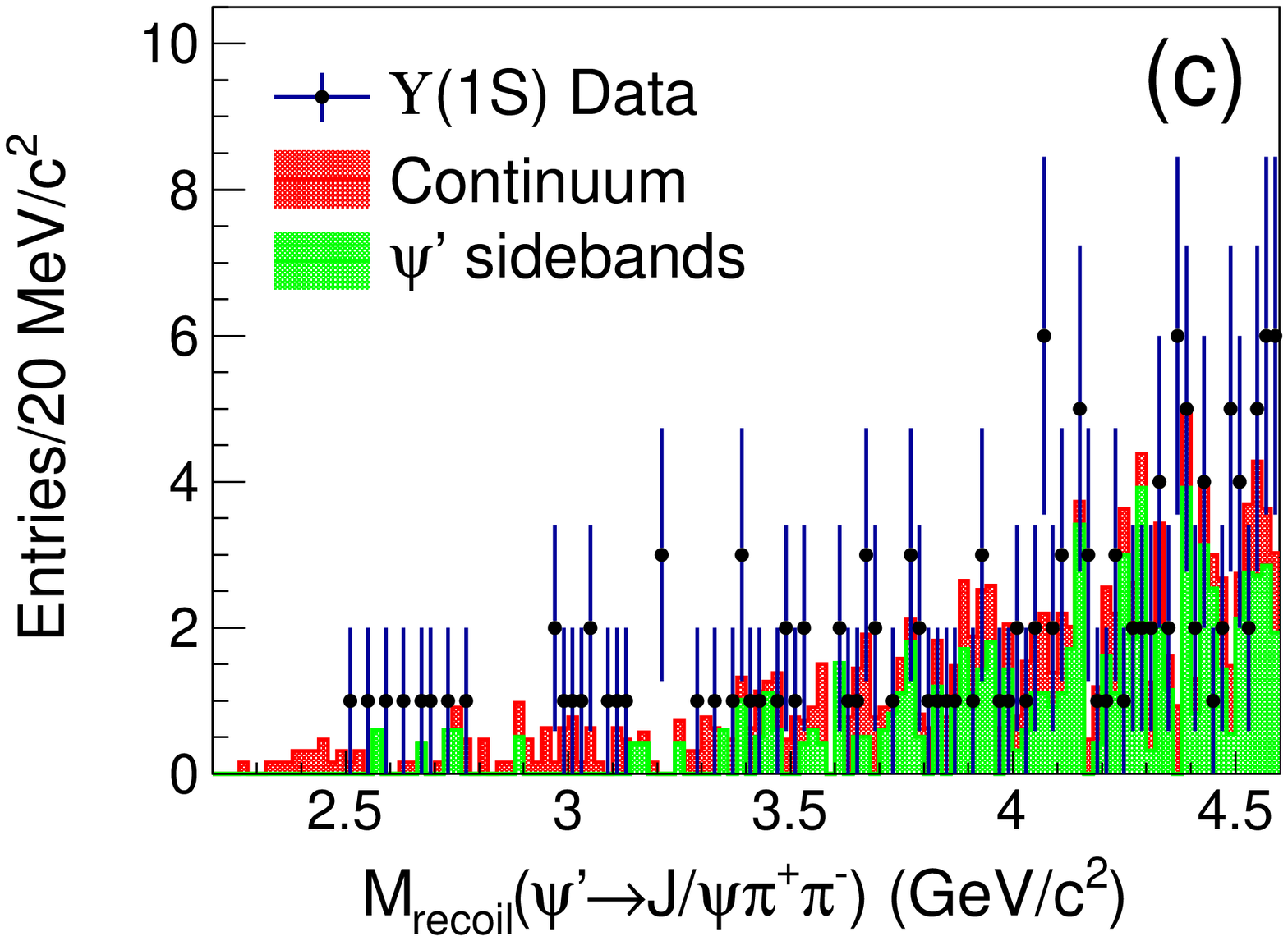}
        \end{minipage}\\
        \begin{minipage}[c]{0.33\textwidth}
        \centering
        \includegraphics[width=0.99\textwidth]{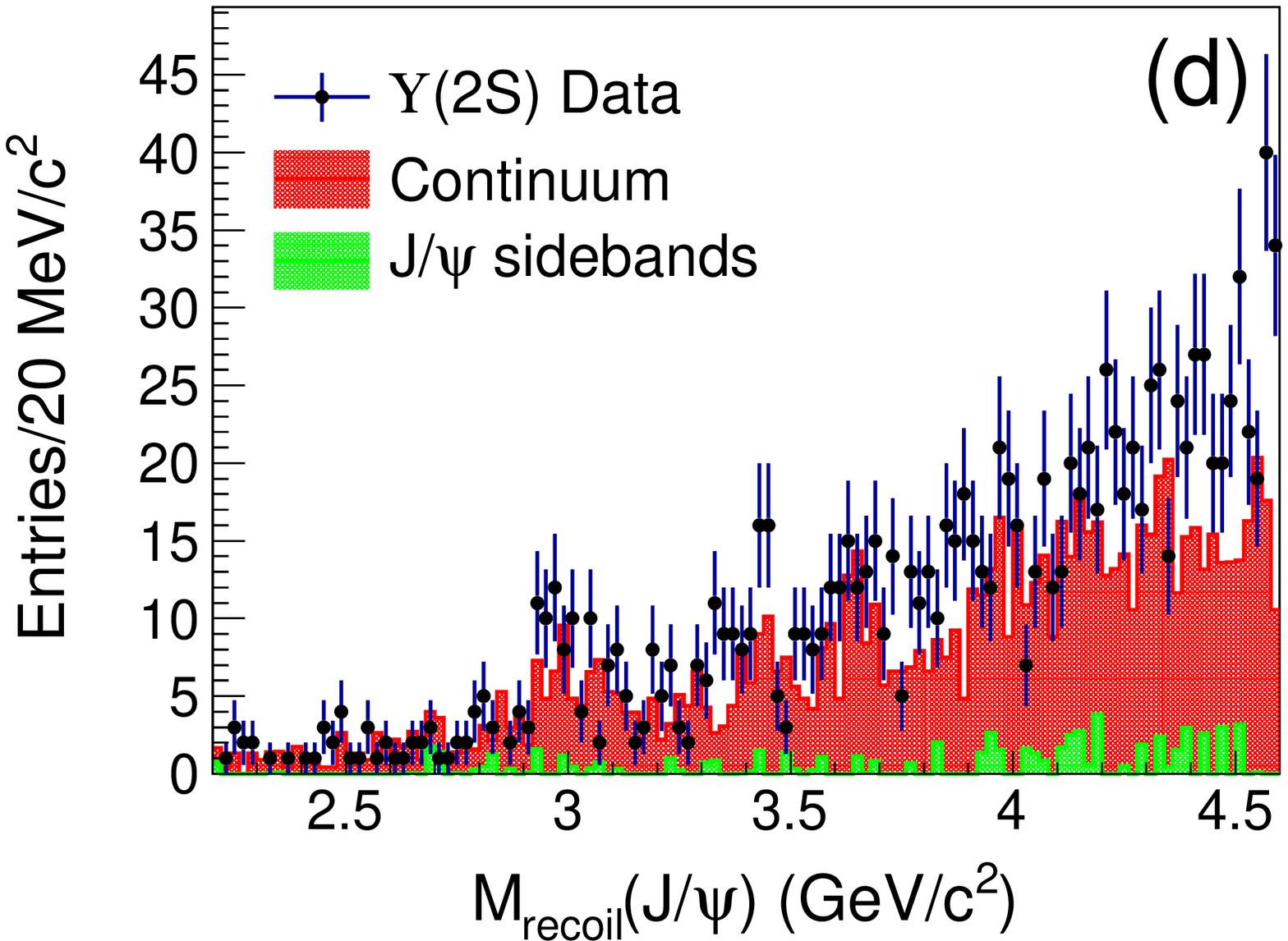}
        \end{minipage}%
        \begin{minipage}[c]{0.33\textwidth}
        \centering
        \includegraphics[width=0.99\textwidth]{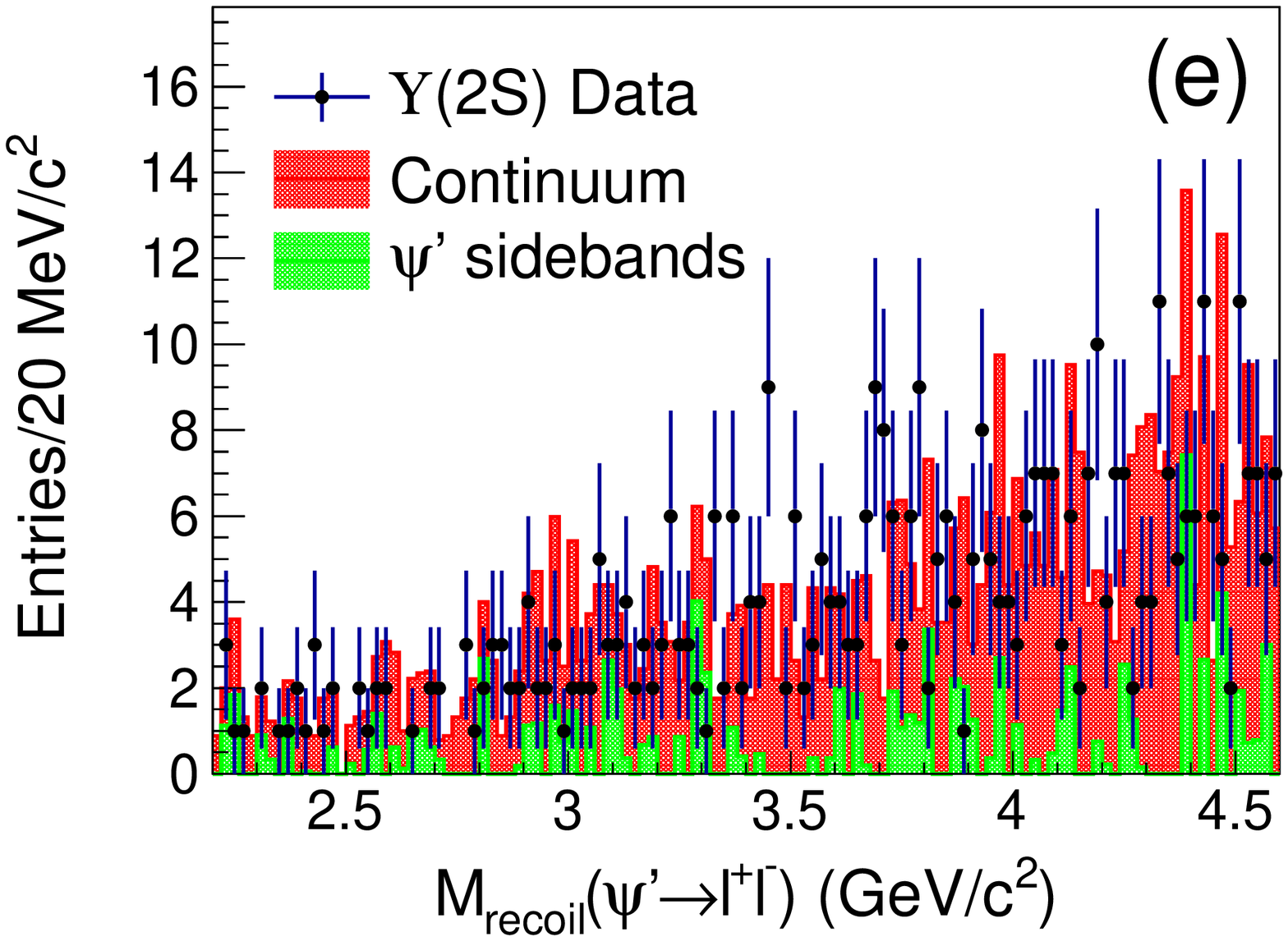}
        \end{minipage}%
        \begin{minipage}[c]{0.33\textwidth}
        \centering
        \includegraphics[width=0.99\textwidth]{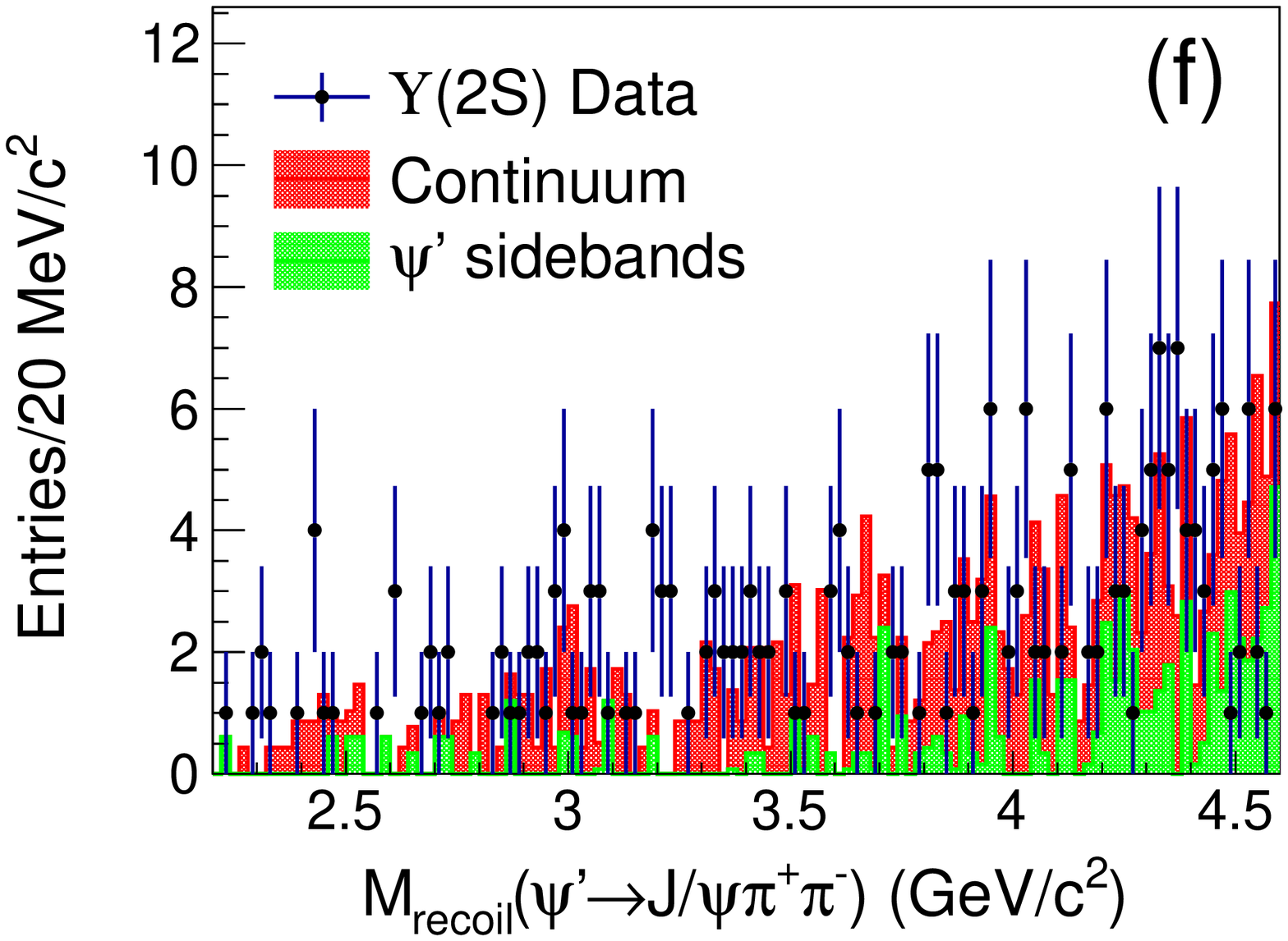}
        \end{minipage}
      \caption{{Distributions of the recoil masses against the reconstructed $J/\psi(\rightarrow \ell^+\ell^-)$,
      $\psi'(\rightarrow \ell^+\ell^-)$, $\psi'(\rightarrow J/\psi\pi^+\pi^-)$
      within the $J/\psi$ or $\psi'$ mass signal regions from left to right.
      The upper and lower three graphs are for $\Upsilon(1S)$ and $\Upsilon(2S)$ decays, respectively.
      The red-shaded histograms are from the normalized continuum sample at $\sqrt{s}=10.52~\mathrm{GeV}$,
      and the green ones represent the scaled $J/\psi(\psi')$ mass sideband backgrounds
      from $\Upsilon$ decays.}}\label{fig-data-Y2S}
    \end{figure}


    After all event selections,
    no peaking background in any charmonium signal region is found from
    the $\Upsilon(1S)$ or $\Upsilon(2S)$ generic MC samples.
    Typical $\Upsilon$ decay samples include three categories:
    $\Upsilon$ decay signal events, $\Upsilon$ decay background events, and continuum events.
    The backgrounds with non-$J/\psi(\psi')$ from $\Upsilon$ decay
    are estimated by normalizing the $J/\psi(\psi')$ mass sideband events to their signal regions.
    The large continuum data sample at $\sqrt{s}=10.52~\mathrm{GeV}$ is used to estimate the continuum contributions in our data samples
    by extrapolating down to the $\Upsilon(1S)$ or $\Upsilon(2S)$ resonance.
    The scale factor of the extrapolation is computed with
    $f_{\mathrm{scale}}=\frac{\mathcal{L}_{\Upsilon}}{\mathcal{L}_{\mathrm{con}}}\frac{\sigma_{\Upsilon}}{\sigma_{\mathrm{con}}}\frac{\varepsilon_{\Upsilon}}{\varepsilon_{\mathrm{con}}}$,
    where $\frac{\mathcal{L}_{\Upsilon}}{\mathcal{L}_{\mathrm{con}}}$, $\frac{\sigma_{\Upsilon}}{\sigma_{\mathrm{con}}}$,
    and $\frac{\varepsilon_{\Upsilon}}{\varepsilon_{\mathrm{con}}}$
    are the ratios of the luminosity, cross sections, and efficiencies, respectively, at the $\Upsilon$ and continuum points.
    For the nominal results, the efficiencies are obtained from MC simulations;
    their ratios in $\Upsilon$ and continuum events
    are equal for all decay modes of $J/\psi(\psi')$
    and the cross sections of the target channels
    are scaled to be proportional to $1/s^4$ ($\sqrt{s}=E_{\mathrm{CM}}$) \cite{PhysRevD.67.054007,PhysRevD.84.034022};
    the corresponding scale factors are about $0.16$ and $0.44$ for $\Upsilon(1S)$ and $\Upsilon(2S)$, respectively.

    Figure~\ref{fig-data-Y2S} shows the distributions of the recoil masses against the
    reconstructed $J/\psi(\rightarrow \ell^+\ell^-)$, $\psi'(\rightarrow \ell^+\ell^-)$,
    and $\psi'(\rightarrow J/\psi\pi^+\pi^-)$ within their signal regions.
    The upper (lower) three graphs are for the $\Upsilon(1S)$ ($\Upsilon(2S)$) decays.
    The green-shaded histograms are the scaled $J/\psi(\psi')$ mass sideband backgrounds
    from $\Upsilon$ decays.
    Contributions from  $e^+e^-$ annihilation with the same final states have been subtracted
    from the sideband distributions to avoid double-counting of continuum events.
    The red-shaded histograms represent the normalized continuum backgrounds,
    whose estimation is described in the previous paragraph.
    In the spectrum of the $J/\psi$ recoil mass in $\Upsilon(1S)$ decays in Fig.~\ref{fig-data-Y2S}(a),
    the sharp peak that appears at $3.51~\mathrm{GeV/{\mathit c}^2}$ is likely to be that of the $\chi_{c1}$,
    for which the width is as narrow as $0.86~\mathrm{MeV/{\mathit c}^2}$~\cite{PhysRevD.86.010001}.
    A slight enhancement around $3.94~\mathrm{GeV/{\mathit c}^2}$ may also be seen;
    no other distinct charmonium signal is observed.
    For the $\psi'\rightarrow \ell^+\ell^-$ mode in Fig.~\ref{fig-data-Y2S}(b)
    and $\psi'\rightarrow J/\psi\pi^+\pi^-$ mode in Fig.~\ref{fig-data-Y2S}(c),
    the $\psi'$ mass sidebands and continuum backgrounds together contribute essentially all the events within the $\psi'$ signal region.
    The $J/\psi$ recoil mass distribution in Fig.~\ref{fig-data-Y2S}(d) reveals weak possible signals around the nominal masses
    of the $\eta_c$, $\chi_{c0}$ and $\eta_c(2S)$ 
    in $\Upsilon(2S)$ decays.
    However, after subtracting the continuum contribution,
    the surviving events are consistent with
    the combinatorial background.
    Similarly to the two $\psi'$ decay modes in $\Upsilon(1S)$ decays,
    only backgrounds are found in the $\psi'$ recoil mass distributions for the $\Upsilon(2S)$ decays,
    as shown in Figs.~\ref{fig-data-Y2S}(e) and \ref{fig-data-Y2S}(f).

    Another background in $\Upsilon(2S)$ decays is the intermediate transition
    $\Upsilon(2S)$ $\to$ $\pi^+$$\pi^-$$\Upsilon(1S)$ or $\pi^0$$\pi^0$$\Upsilon(1S)$
    with $\Upsilon(1S)$ decaying into double charmonia.
    Such contamination is examined with the recoil masses
    of additionally selected $\pi^+\pi^-$ or $\pi^0\pi^0$ pairs to check for $\Upsilon(1S)$ signals.
    After all event selections, the ratios of such backgrounds are $(9.6\pm1.7)\%$ and
    $(15.0\pm2.8)\%$ for the $J/\psi+X$ and $\psi'+X$ processes, respectively,
    by fitting the recoil mass spectra of $\pi^0\pi^0$ and $\pi^+\pi^-$ pairs.
    However, the corresponding distribution of the mass recoiling
    against the $J/\psi(\psi')$ is smooth;
    therefore, the contamination is non-peaking.
    Here, a $\pi^0$ candidate is reconstructed
    from a pair of good photons~\cite{BN1165} with an invariant mass within
    $15~\mathrm{MeV/{\mathit c}^2}$ of the $\pi^0$ nominal mass.
    We require $\chi^2<20$, where $\chi^2$ is from the mass-constrained fit of $\pi^0\rightarrow\gamma\gamma$.


    An unbinned extended simultaneous likelihood fit is applied to
    the spectra of the mass recoiling against the $J/\psi$ or $\psi'$
    to extract the signal yields in the $\Upsilon(1S,2S)$ and continuum data samples.
    For $\psi'+X$ processes, the decay modes $\psi'\rightarrow \ell^+\ell^-$
    and $\psi'\rightarrow J/\psi\pi^+\pi^-$ are treated together
    to obtain the total yield of every $\psi'$ recoil $c\bar{c}$ signal.
    That is to say, in the fit to the $\psi'$ recoil mass spectra,
    in addition to the simultaneous fit applied to the $\Upsilon$ and continuum data samples,
    we also apply a simultaneous fit to these two $\psi'$ decay modes.
    The ratio of any charmonium-like yields between the $\psi'\rightarrow \ell^+\ell^-$ and $\psi'\rightarrow J/\psi\pi^+\pi^-$ modes
    is fixed to the ratio of the MC-determined efficiencies between these two $\psi'$ decay modes
    with all the intermediate-state branching fractions included.

    The signal shapes of all the recoil $c\bar{c}$ states are determined from MC simulations
    with the mass resolutions of $31~\mathrm{MeV}/c^2$, $24~\mathrm{MeV}/c^2$,
    $23~\mathrm{MeV}/c^2$, $19~\mathrm{MeV}/c^2$, and $18~\mathrm{MeV}/c^2$
    for the recoiling $\eta_c$, $\chi_{cJ}$, $\eta_c(2S)$, $X(3940)$, and $X(4160)$, respectively.
    In the MC simulations, for all the recoil $c\bar{c}$ states,
    the world-average resonance parameters are used with masses fixed at
    2.984 GeV/$c^2$, 3.097 GeV/$c^2$, 3.415 GeV/$c^2$, 3.511 GeV/$c^2$,
    3.556 GeV/$c^2$, 3.639 GeV/$c^2$, 3.942 GeV/$c^2$ and 4.156 GeV/$c^2$ for $\eta_c$, $J/\psi$,
    $\chi_{c0}$, $\chi_{c1}$, $\chi_{c2}$, $\eta_c(2S)$, $X(3940)$, and $X(4160)$, respectively~\cite{PhysRevD.86.010001}.
    Because of the production-channel dependence of the transition matrix element
    for a description of the $\eta_c$ line shape \cite{PhysRevLett.108.222002},
    a smearing Gaussian function with free parameters is introduced there to
    improve the fit accuracy and to account for possible discrepancies between data and MC .
    In other words, the $\eta_c$ shape is described with the MC-determined shape
    convolved with this Gaussian function.
    The other $c\bar{c}$ signals are described directly by the MC-determined shapes.
    In the fit to the $\Upsilon$ candidates,
    a Chebychev polynomial background shape is used for the $\Upsilon(1S,2S)$
    decay backgrounds in addition to the normalized continuum contribution.
    Since the fit range includes
    the region over the $D\bar{D}$ threshold ($\approx3.73~\mathrm{GeV/{\mathit c}^2}$),
    a threshold term  proportional to $(M_{\mathrm{recoil}}(c\bar{c})-2m_D)^n$ is added, where $n$ is a free parameter
    and $m_D$ is the $D$ meson nominal mass.
    This term is added in the background parametrization with a
    free normalization to account for the possible contribution from
    $\Upsilon(1S,2S)/e^+e^-\rightarrow J/\psi(\psi')D^{(\ast)}\bar{D}^{(\ast)}$ .

\begin{figure}[!htb]
      \centering
        \begin{minipage}[c]{0.33\textwidth}
        \centering
        \includegraphics[width=0.99\textwidth]{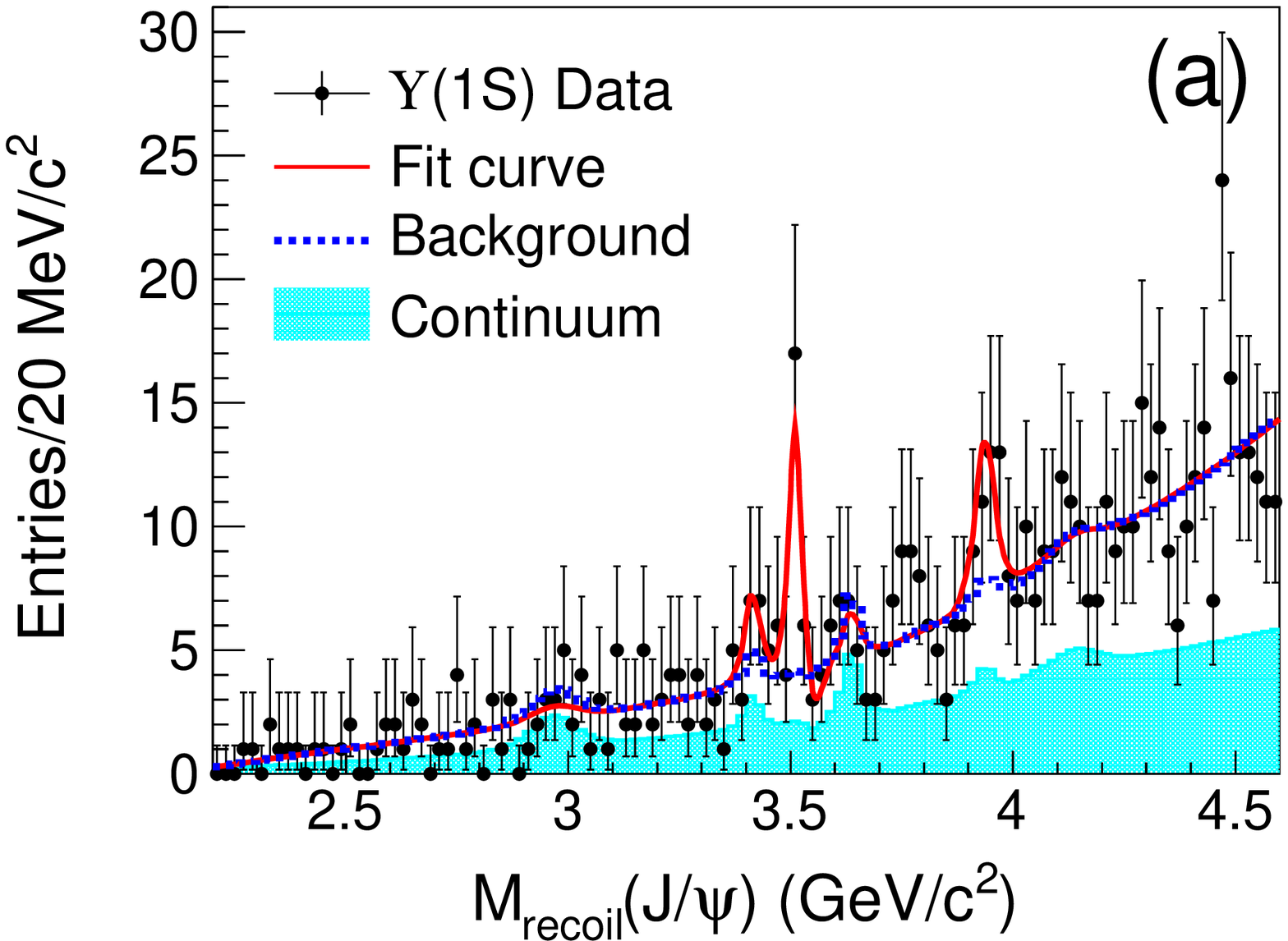}
        \end{minipage}%
        \begin{minipage}[c]{0.33\textwidth}
        \centering
        \includegraphics[width=0.99\textwidth]{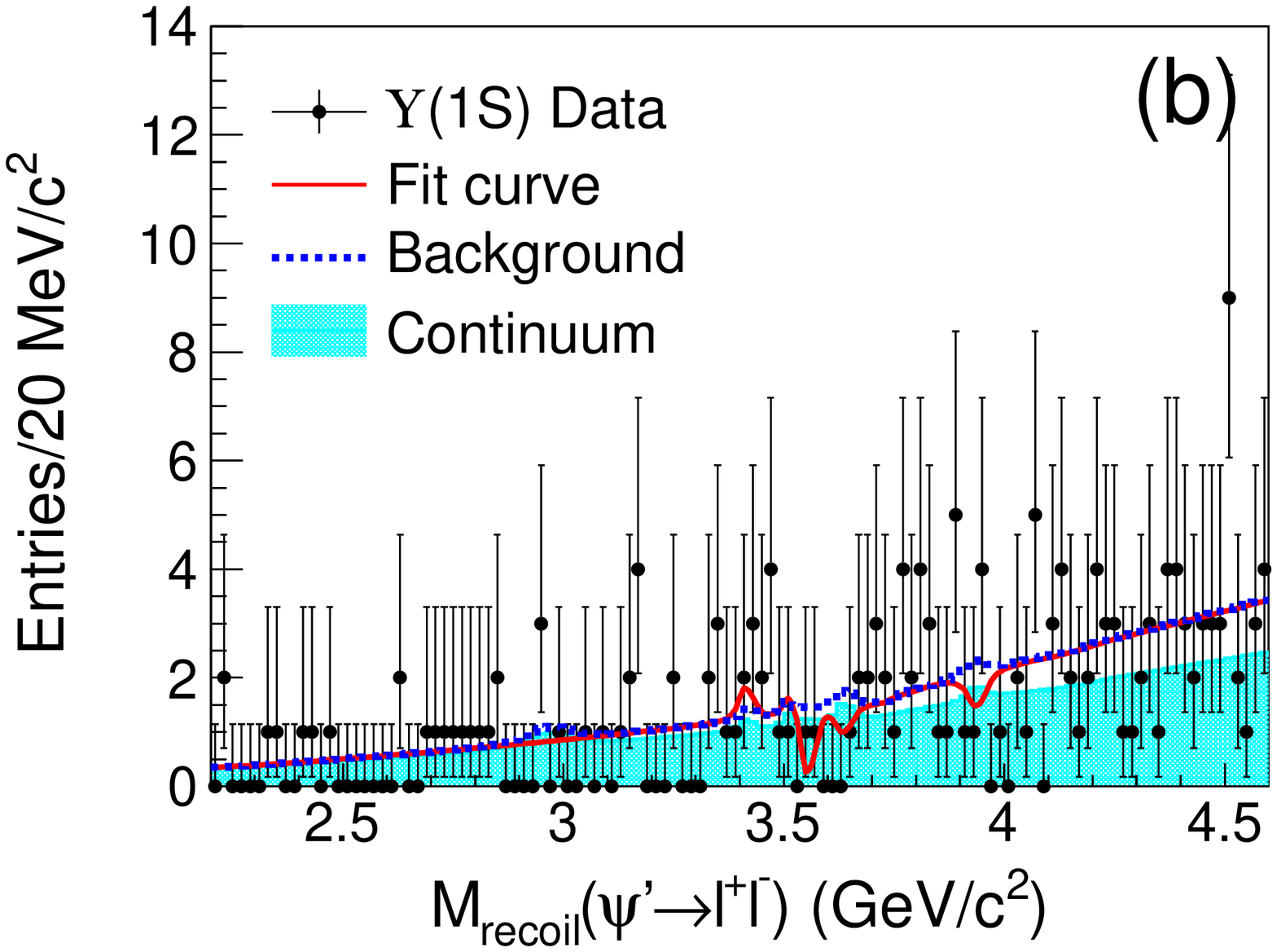}
        \end{minipage}%
        \begin{minipage}[c]{0.33\textwidth}
        \centering
        \includegraphics[width=0.99\textwidth]{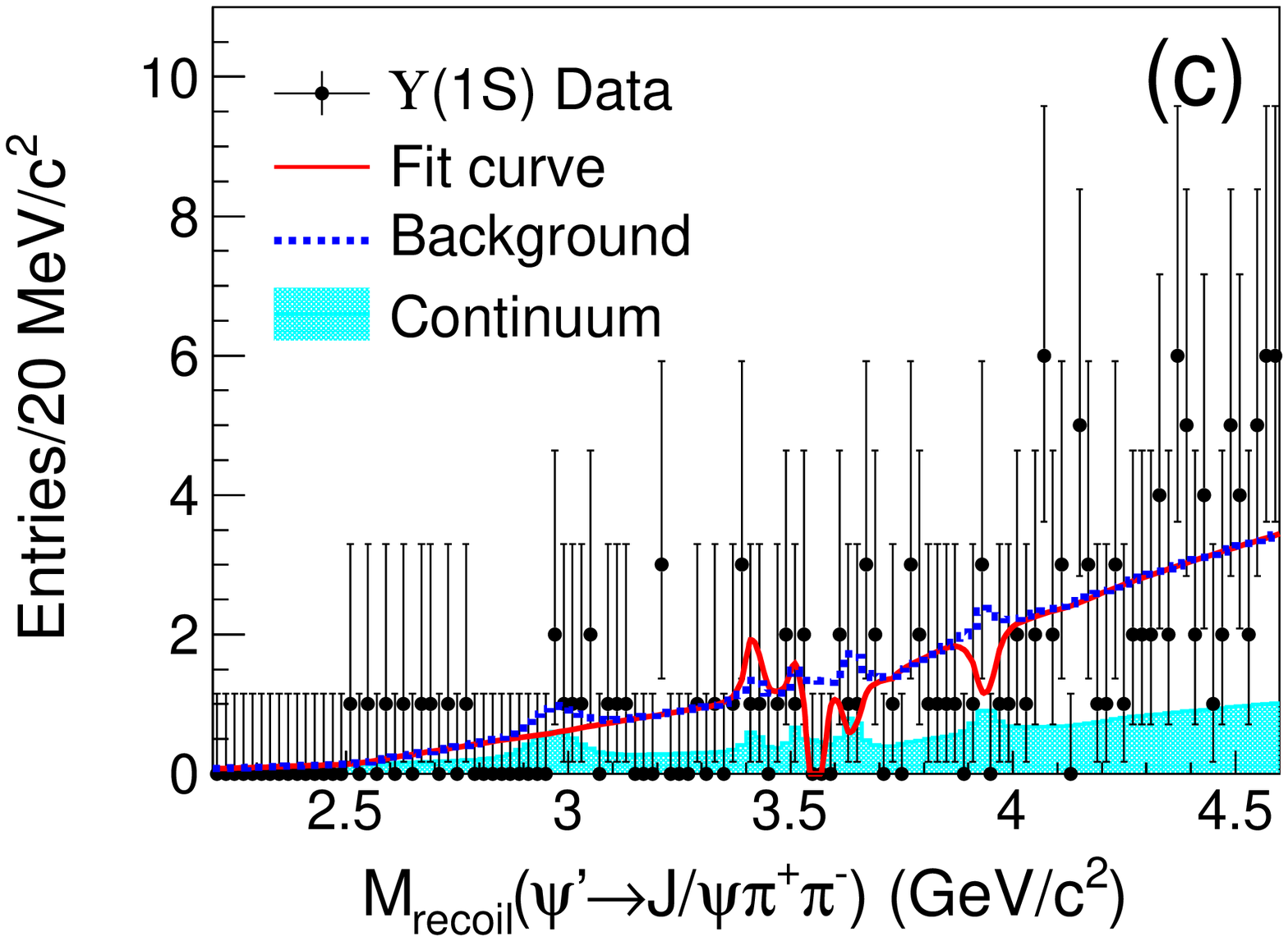}
        \end{minipage}\\
      \caption{{Fit to the recoil mass spectra against the
      (a) $J/\psi(\rightarrow \ell^+\ell^-)$, (b) $\psi'(\rightarrow \ell^+\ell^-)$ and (c) $\psi'(\rightarrow J/\psi\pi^+\pi^-)$
      in $\Upsilon(1S)$ decays from data (points with error bars).
      The red solid curves are the nominal fits
      and the blue-dashed curves show the total background.
      The fitted normalized continuum contributions are represented by the cyan-shaded histograms.}}\label{fig-data-fit-Y1S}
    \end{figure}

\begin{figure}[!htb]
      \centering
        \begin{minipage}[c]{0.33\textwidth}
        \centering
        \includegraphics[width=0.99\textwidth]{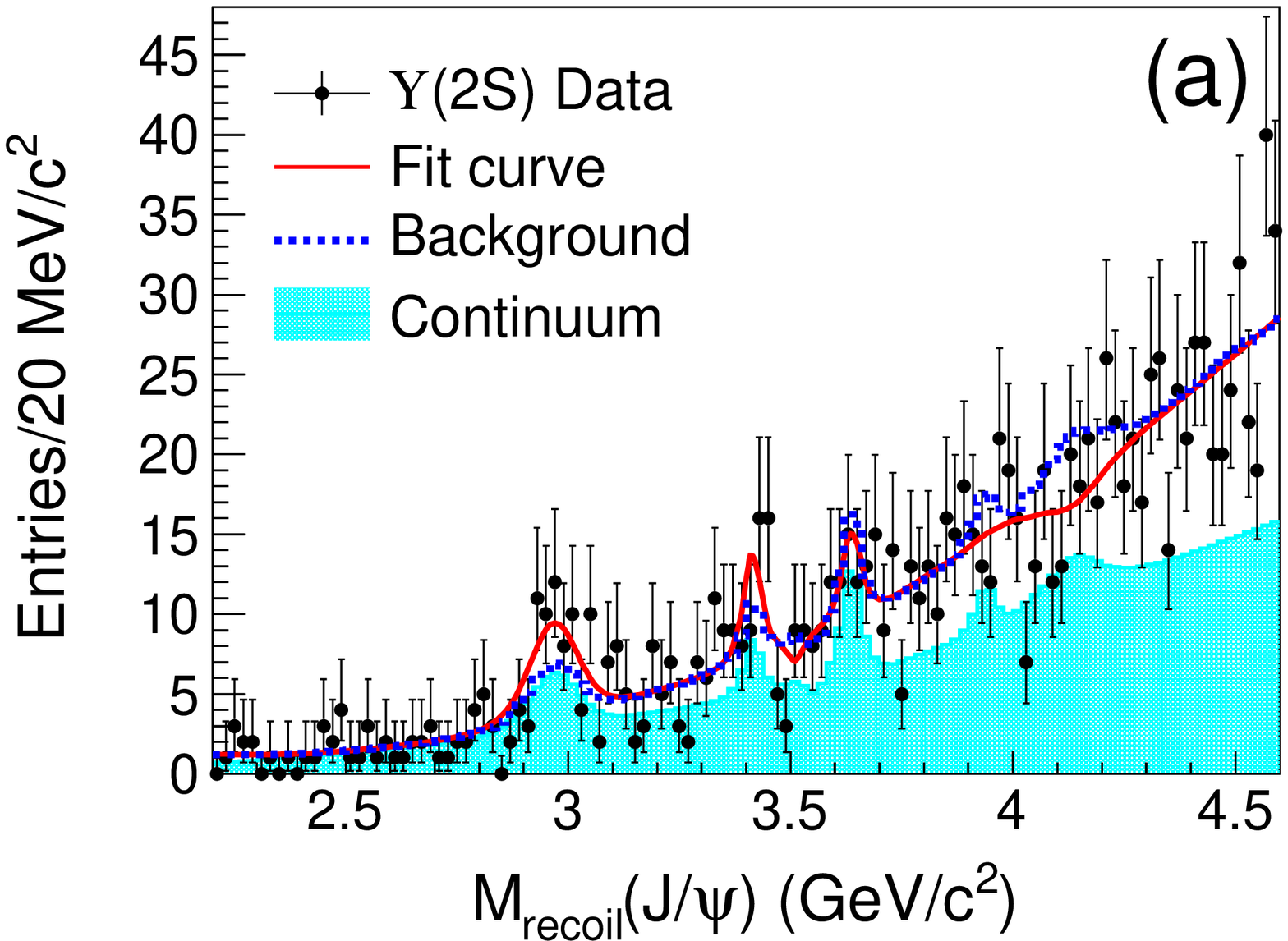}
        \end{minipage}%
        \begin{minipage}[c]{0.33\textwidth}
        \centering
        \includegraphics[width=0.99\textwidth]{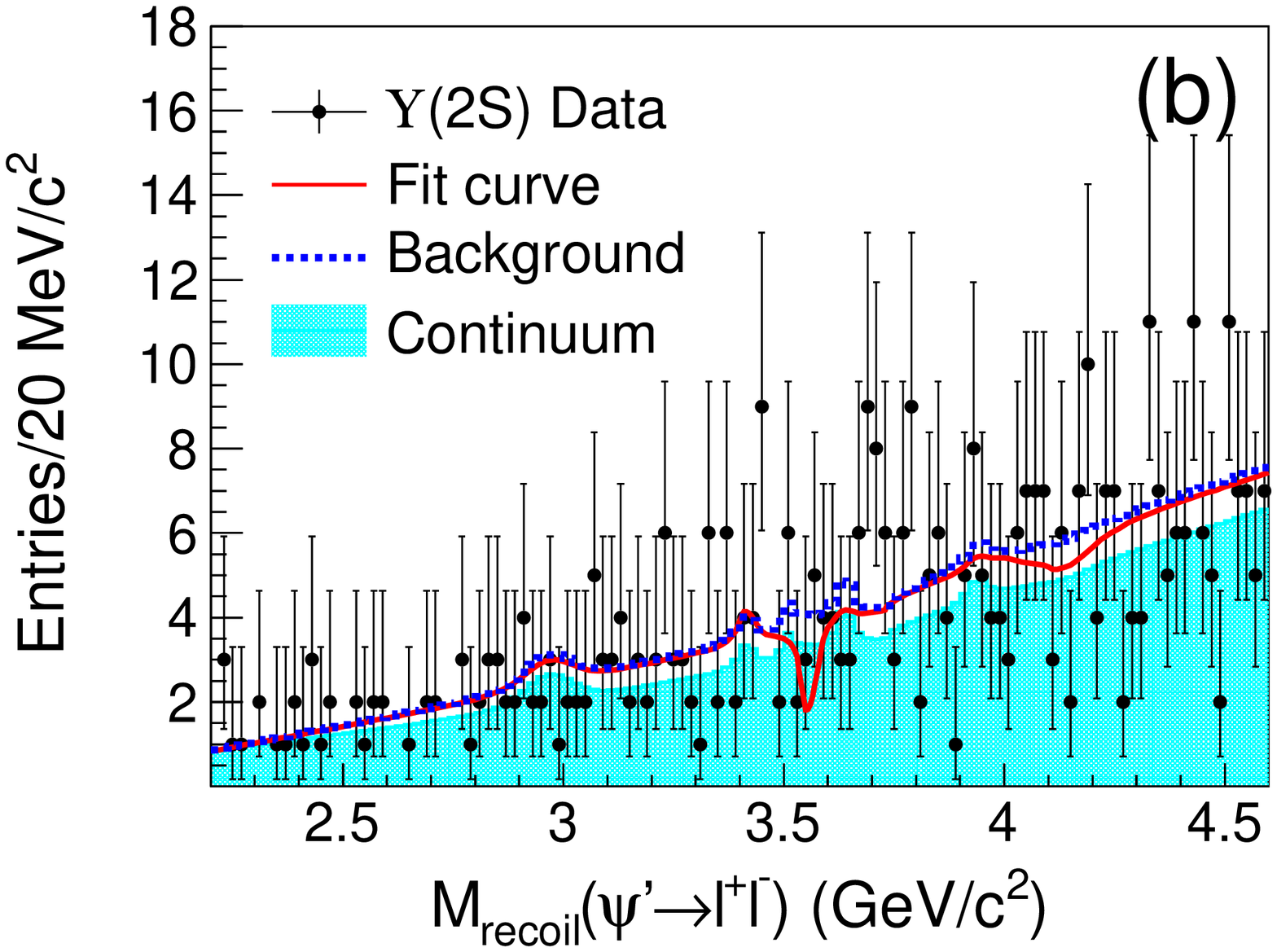}
        \end{minipage}%
        \begin{minipage}[c]{0.33\textwidth}
        \centering
        \includegraphics[width=0.99\textwidth]{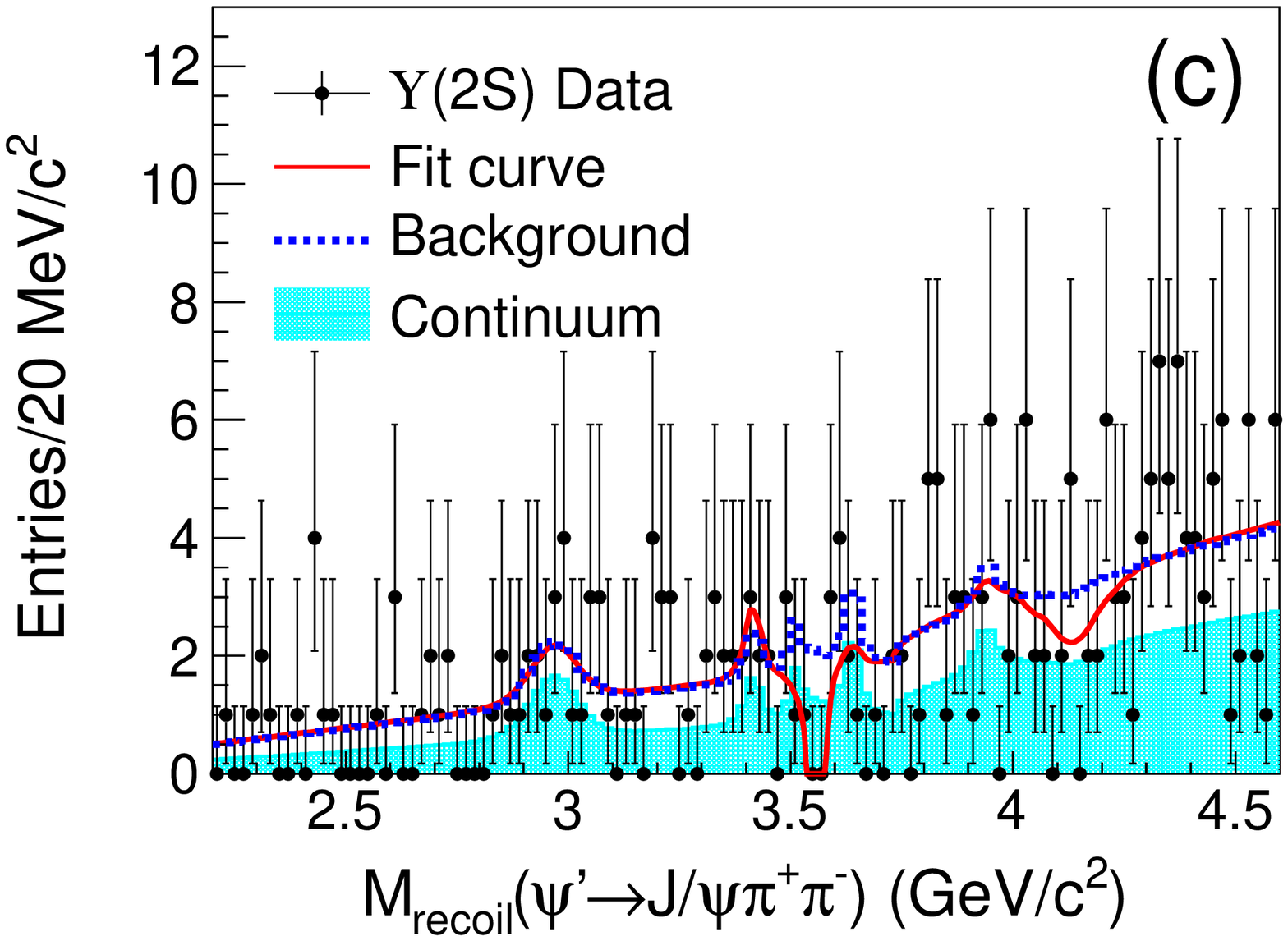}
        \end{minipage}\\
      \caption{{Fit to the recoil mass spectra against
      (a) $J/\psi(\rightarrow \ell^+\ell^-)$, (b) $\psi'(\rightarrow \ell^+\ell^-)$ and (c) $\psi'(\rightarrow J/\psi\pi^+\pi^-)$
      in $\Upsilon(2S)$ decays from data (points with error bars).
      The representations of the curves and histograms in each graph
      match those in Fig.~\ref{fig-data-fit-Y1S}.
      }}\label{fig-data-fit-Y2S}
    \end{figure}

    The fit range and results to the spectra of the recoil mass against $J/\psi$ and $\psi'$
    are shown in Figs.~\ref{fig-data-fit-Y1S} and~\ref{fig-data-fit-Y2S}
    from the $\Upsilon(1S)$ and $\Upsilon(2S)$ data samples, respectively.
    The points with error bars represent the $\Upsilon(1S,2S)$ events.
    The red solid curves give the nominal fit results while the
    blue-dashed curves are the estimated total background.
    The cyan-shaded histograms are the fitted normalized continuum contributions
    under the $J/\psi(\psi')$ signal region.
    The fitted signal yields ($N_{\rm fit}$) of every recoil charmonium state are listed in Table~\ref{Table-conclusion}.

    \begin{table*}[!htpb]
        \caption{\label{Table-conclusion} Results of the search for $\Upsilon(1S)$ and $\Upsilon(2S)$ decays into double charmonia.
        For each decay mode, $N_{fit}$ represents the number of fitted signal events,
        $N_{\rm up}$ is the upper limit on the number of signal events,
        $\varepsilon$ is the reconstruction efficiency,
        $\sigma_{syst}$ is the total systematic error,
        $\Sigma$ is the signal significance with systematic error included,
        $\mathcal{B}_{R}$ is the measured branching fraction where the upper limit is at $90\%$ C.L.,
        and $\mathcal{B}_{\rm th}$ is taken from the theoretical predictions~\cite{PhysRevD.76.074007,PhysRevD.87.094004}.}
        \begin{ruledtabular}
        \begin{tabular}{lrrcrrrr}
         Channels & $N_{fit}\ \ \ $ & $N_{\rm up}$ & $\varepsilon(\%)$ & $\sigma_{syst}(\%)$ & $\Sigma~(\sigma) $ & $\mathcal{B}_{R}(\times10^{-6})$ & $\mathcal{B}_{\rm th}(\times10^{-6})$\\
        \hline
        $\Upsilon(1S)\rightarrow J/\psi+\eta_c$    & $-4.9\pm6.3$  & $8.1$  & $3.71$ & $8.1$  & $-$   & $<2.2$ & $3.9^{+5.6}_{-2.3}$ \\
        $\Upsilon(1S)\rightarrow J/\psi+\chi_{c0}$ & $6.0\pm5.6$   & $14.4$ & $4.25$ & $5.1$  & $1.3$ & $<3.4$ & $1.3$ \\
        $\Upsilon(1S)\rightarrow J/\psi+\chi_{c1}$ & $19.9\pm6.2$  & $-$    & $4.98$ & $5.9$  & $4.6$ & $3.90\pm1.21\pm0.23$ & $4.9$ \\
        $\Upsilon(1S)\rightarrow J/\psi+\chi_{c2}$ & $-3.2\pm4.0$  & $6.4$  & $4.71$ & $4.7$  & $-$   & $<1.4$& $0.20$ \\
        $\Upsilon(1S)\rightarrow J/\psi+\eta_c(2S)$& $-2.2\pm6.0$  & $9.3$  & $4.32$ & $5.2$  & $-$   & $<2.2$ & $2.0^{+3.4}_{-1.4}$ \\
        $\Upsilon(1S)\rightarrow J/\psi+X(3940)$   & $18.4\pm8.8$  & $30.9$ & $5.67$ & $8.4$  & $2.6$ & $<5.4$ & $-$ \\
        $\Upsilon(1S)\rightarrow J/\psi+X(4160)$   & $-0.7\pm15.0$ & $22.7$ & $5.28$ & $19.7$ & $-$   & $<5.4$ & $-$ \\
        $\Upsilon(1S)\rightarrow \psi'+\eta_c$     & $-4.6\pm4.0$  & $5.8$  & $1.58$ & $13.5$ & $-$   & $<3.6$ & $1.7^{+2.4}_{-1.0}$ \\
        $\Upsilon(1S)\rightarrow \psi'+\chi_{c0}$  & $2.5\pm4.2$   & $10.6$ & $1.60$ & $17.7$ & $0.7$ & $<6.5$ & $-$ \\
        $\Upsilon(1S)\rightarrow \psi'+\chi_{c1}$  & $0.6\pm3.7$   & $7.9$  & $1.68$ & $21.5$ & $0.2$ & $<4.5$ & $-$ \\
        $\Upsilon(1S)\rightarrow \psi'+\chi_{c2}$  & $-6.5\pm2.4$  & $3.5$  & $1.64$ & $7.1$  & $-$   & $<2.1$ &  $-$ \\
        $\Upsilon(1S)\rightarrow \psi'+\eta_c(2S)$ & $-5.4\pm3.6$  & $5.3$  & $1.68$ & $20.5$ & $-$   & $<3.2$ & $0.8^{+1.4}_{-0.6}$ \\
        $\Upsilon(1S)\rightarrow \psi'+X(3940)$    & $-6.7\pm4.0$  & $5.6$  & $1.92$ & $11.8$ & $-$   & $<2.9$ & $-$ \\
        $\Upsilon(1S)\rightarrow \psi'+X(4160)$    & $-0.3\pm10.3$ & $17.2$ & $1.86$ & $21.8$ & $-$   & $<2.9$ & $-$ \\
        $\Upsilon(2S)\rightarrow J/\psi+\eta_c$    & $18.8\pm11.8$ & $35.7$ & $3.61$ & $16.9$ & $2.2$ & $<5.4$ & $2.6^{+3.7}_{-1.6}$ \\
        $\Upsilon(2S)\rightarrow J/\psi+\chi_{c0}$ & $9.3\pm9.4$   & $21.5$ & $4.17$ & $6.4$  & $1.3$ & $<3.4$ & $1.1$ \\
        $\Upsilon(2S)\rightarrow J/\psi+\chi_{c1}$ & $-4.0\pm6.5$  & $8.4$  & $4.95$ & $5.8$  & $-$   & $<1.2$ & $4.1$ \\
        $\Upsilon(2S)\rightarrow J/\psi+\chi_{c2}$ & $2.3\pm7.4$   & $13.1$ & $4.57$ & $6.8$  & $0$   & $<2.0$ & $0.17$ \\
        $\Upsilon(2S)\rightarrow J/\psi+\eta_c(2S)$& $-4.7\pm10.8$ & $13.7$ & $4.23$ & $10.4$ & $-$   & $<2.5$ & $1.3^{+2.1}_{-0.9}$ \\
        $\Upsilon(2S)\rightarrow J/\psi+X(3940)$   & $-8.8\pm11.9$ & $14.0$ & $5.65$ & $16.3$ & $-$   & $<2.0$ & $-$ \\
        $\Upsilon(2S)\rightarrow J/\psi+X(4160)$   & $-40.3\pm22.2$& $14.9$ & $5.37$ & $18.6$ & $-$   & $<2.0$ & $-$ \\
        $\Upsilon(2S)\rightarrow \psi'+\eta_c$     & $-1.4\pm8.4$  & $11.9$ & $1.56$ & $8.6$  & $-$   & $<5.1$ & $1.1^{+1.6}_{-0.7}$ \\
        $\Upsilon(2S)\rightarrow \psi'+\chi_{c0}$  & $1.6\pm6.1$   & $11.3$ & $1.63$ & $8.2$  & $0.3$ & $<4.7$ & $-$ \\
        $\Upsilon(2S)\rightarrow \psi'+\chi_{c1}$  & $-3.7\pm4.5$  & $6.2$  & $1.66$ & $6.9$  & $-$   & $<2.5$ & $-$ \\
        $\Upsilon(2S)\rightarrow \psi'+\chi_{c2}$  & $-13.5\pm5.2$ & $4.9$  & $1.66$ & $6.9$  & $-$   & $<1.9$ & $-$ \\
        $\Upsilon(2S)\rightarrow \psi'+\eta_c(2S)$ & $-5.0\pm6.6$  & $8.0$  & $1.66$ & $7.7$  & $-$   & $<3.3$ & $0.5^{+0.9}_{-0.4}$ \\
        $\Upsilon(2S)\rightarrow \psi'+X(3940)$    & $-2.0\pm7.3$  & $10.7$ & $1.96$ & $7.9$  & $-$   & $<3.9$ & $-$ \\
        $\Upsilon(2S)\rightarrow \psi'+X(4160)$    & $-13.1\pm14.0$& $12.4$ & $1.89$ & $10.9$ & $-$   & $<3.9$ & $-$ \\
        \end{tabular}
        \end{ruledtabular}
    \end{table*}


    Several sources of systematic errors are taken into account in the branching fraction measurements.
    Tracking efficiency uncertainty is estimated to be
    0.35\% per track with high momentum and is additive.
    Based on the measurements of the identification efficiencies of
    the lepton pair with $\gamma\gamma\rightarrow \ell^+\ell^-$  and the pion using the $D^\ast$ sample,
    the MC simulates data with uncertainties within about $1.8\%$ and $1.3\%$ for each lepton and pion, respectively.
    As the trigger efficiency evaluated from a trigger simulation
     is greater than $99.9\%$,
    its uncertainty can be neglected.
    The errors on the branching fractions of the intermediate states
    are taken from the Particle Data Group \cite{PhysRevD.86.010001},
    which are about $1.1\%$, $6.3\%$ and $1.2\%$
    for $J/\psi\rightarrow \ell^+\ell^-$, $\psi'\rightarrow \ell^+\ell^-$ and $\psi'\rightarrow J/\psi\pi^+\pi^-$, respectively;
    the weighted average for the two $\psi'$ decay modes is about $3.5\%$.
    For the charmonium states with generic decays,
    the unknown decay channels are generated by the Lund fragmentation model in PYTHIA \cite{JHEP2006.026}.
    By generating different sets of MC samples with different relative probabilities
    to produce the various possible $q\bar{q}$ ($q=u,~d,~s$) pairs,
    the largest difference in the efficiencies is found to be less than $0.1\%$ and thus is neglected.
    The uncertainty due to the $N_{\rm {ch}}>4$ requirement is at the 1.0\% level, determined by changing
    the known decay branching fractions of recoil charmonium states to the final states
    with $N_{\rm {ch}}<5$ by 1$\sigma$~\cite{PhysRevD.86.010001}.
    By varying the background shapes or the order of the Chebychev polynomial,
    as well as the fitted range and the width of the smearing Gaussian within $\pm1\sigma$,
    the deviation of the upper limits on the number of the signal events is found
    to be between $2.0\%$ and $24.1\%$, depending on the decay mode.
    The MC statistical errors are estimated using the reconstruction efficiencies and the number of generated events,
    which are at most  $1.8\%$.
    The uncertainties associated with the total number of $\Upsilon(1S)$ and $\Upsilon(2S)$ events are $2.0\%$ and $2.3\%$, respectively.
    Assuming that all the sources are independent and summed in quadrature,
    the total systematic errors ($\sigma_{\rm syst}$) are evaluated and listed in Table~\ref{Table-conclusion}.

    Since few distinct signals are observed,
    the upper limit on the number of signal events ($N_{\rm up}$) 
    is determined at the $90\%$ confidence level (C.L.)
    by solving the equation $\int^{N_{\rm up}}_0\mathcal{L}(x)dx/\int^{+\infty}_0\mathcal{L}(x)dx$$=0.9$,
    where $x$ is the number of fitted signal events
    and $\mathcal{L}(x)$ is the likelihood function in the fit to the data,
    convolved here with a Gaussian function whose width equals the total systematic uncertainty.
    The value of $N_{\rm up}$ for each mode, which requires the signal yields to be non-negative in the fit,
    is listed in Table~\ref{Table-conclusion}  along with the corresponding calculated branching fraction ($\BR_{R}$) or its upper limit.
    The theoretical predictions ($\mathcal{B}_{\rm th}$) from Refs. \cite{PhysRevD.76.074007,PhysRevD.87.094004} are also tabulated.
    Due to the sensitivity to the choices of some parameters
    such as the charm-quark mass ($m_c$), NRQCD matrix elements, and QCD coupling constant ($\alpha_s$),
    the central values of $\mathcal{B}_{\rm th}$ have large uncertainties.
    Table~\ref{Table-conclusion} also lists the reconstruction efficiency ($\varepsilon$)
    and the signal significance ($\Sigma$) that is obtained by calculating $\sqrt{-2\ln(\mathcal{L}_0/\mathcal{L}_{\rm max})}$,
    where $\mathcal{L}_0$ and $\mathcal{L}_{\rm max}$ are the likelihoods of the fits without and with a signal component, respectively.
    Here, for the likelihood function the width of the convolved Gaussian equals the systematic uncertainty
    related to signal yield instead of the total systematic uncertainty.


   To summarize, we have performed a first experimental investigation into
   double-charmonium production in $\Upsilon(1S,2S)$ decays
   by using the Belle data samples of $102\times10^6$ $\Upsilon(1S)$ and $158\times10^6$ $\Upsilon(2S)$ events.
   The evidence for the mode $\Upsilon(1S)\rightarrow J/\psi+\chi_{c1}$ is found,
   for which the branching fraction is measured to be
   $\mathcal{B}(\Upsilon(1S)\rightarrow J/\psi+\chi_{c1})=(3.90\pm1.21 (\rm stat.)\pm0.23 (\rm syst.))\times10^{-6}$
   ($<5.7\times 10^{-6}$ at 90\% C.L.)
   with a signal significance of $4.6\sigma$.
   The 90\% C.L. upper limits are set on the branching fractions of
   the other decays of $\Upsilon(1S,2S)$ into double-charmonium states that
   have a signal significance of less than $3\sigma$.
   Our results are found to be consistent with the theoretical calculations made
   using the NRQCD factorization approach~\cite{PhysRevD.76.074007,PhysRevD.87.094004}.



We thank the KEKB group for the excellent operation of the
accelerator; the KEK cryogenics group for the efficient
operation of the solenoid; and the KEK computer group,
the National Institute of Informatics, and the
PNNL/EMSL computing group for valuable computing
and SINET4 network support.  We acknowledge support from
the Ministry of Education, Culture, Sports, Science, and
Technology (MEXT) of Japan, the Japan Society for the
Promotion of Science (JSPS), and the Tau-Lepton Physics
Research Center of Nagoya University;
the Australian Research Council and the Australian
Department of Industry, Innovation, Science and Research;
Austrian Science Fund under Grant No.~P 22742-N16 and P 26794-N20;
the National Natural Science Foundation of China under Contracts
No.~10575109, No.~10775142, No.~10825524, No.~10875115, No.~10935008
and No.~11175187, the Fundamental Research Funds for the Central
Universities YWF-14-WLXY-013
and CAS center for Excellence in Particle Physics (China);
the Ministry of Education, Youth and Sports of the Czech
Republic under Contract No.~LG14034;
the Carl Zeiss Foundation, the Deutsche Forschungsgemeinschaft
and the VolkswagenStiftung;
the Department of Science and Technology of India;
the Istituto Nazionale di Fisica Nucleare of Italy;
National Research Foundation of Korea Grants
No.~2011-0029457, No.~2012-0008143, No.~2012R1A1A2008330,
No.~2013R1A1A3007772, No.~2014R1A2A2A01005286, No.~2014R1A2A2A01002734,
No.~2014R1A1A2006456;
the BRL program under NRF Grant No.~KRF-2011-0020333, No.~KRF-2011-0021196,
Center for Korean J-PARC Users, No.~NRF-2013K1A3A7A06056592; the BK21
Plus program and the GSDC of the Korea Institute of Science and
Technology Information;
the Polish Ministry of Science and Higher Education and
the National Science Center;
the Ministry of Education and Science of the Russian
Federation and the Russian Federal Agency for Atomic Energy;
the Slovenian Research Agency;
the Basque Foundation for Science (IKERBASQUE) and the UPV/EHU under
program UFI 11/55;
the Swiss National Science Foundation; the National Science Council
and the Ministry of Education of Taiwan; and the U.S.\
Department of Energy and the National Science Foundation.
This work is supported by a Grant-in-Aid from MEXT for
Science Research in a Priority Area (``New Development of
Flavor Physics'') and from JSPS for Creative Scientific
Research (``Evolution of Tau-lepton Physics'').

\bibliography{draftRef}

\begin{thebibliography}{48}
\bibitem [{BN1()}]{BN1165}%
  \BibitemOpen
  \href@noop {} {}\bibinfo {note} {A neutral cluster in the electromagnetic
  calorimeter is reconstructed as a photon if it does not match the
  extrapolated position of any charged track and its energy is greater than 40
  MeV.}\BibitemShut {Stop}%
\end{thebibliography}
%
%
%
%

\end{document}